\pdfoutput=1

\documentclass[11pt]{article}

\usepackage[final]{acl}

\usepackage{times}
\usepackage{latexsym}
\usepackage[T1]{fontenc}
\usepackage[utf8]{inputenc}
\usepackage{microtype}
\usepackage{multirow}
\usepackage{booktabs}
\usepackage{array}
\usepackage{inconsolata}
\usepackage{graphicx}
\usepackage{amsmath}
\usepackage{amssymb}
\usepackage{xcolor}
\usepackage{comment}
\usepackage{xspace}
\usepackage{enumitem}
\usepackage{adjustbox}
\usepackage{bbm}
\usepackage{dsfont}
\usepackage{hyperref}
\usepackage{pdfpages}
\usepackage{caption}
\usepackage{subcaption}
\usepackage{threeparttable}
\usepackage[ruled,vlined]{algorithm2e}

\usepackage{tcolorbox}
\definecolor{promptgray}{gray}{0.97}
\definecolor{framegray}{gray}{0.40}
\usepackage{xcolor}

\newtcolorbox{promptbox}[1]{
     colback=promptgray,
     colframe=framegray,
     fonttitle=\bfseries\small,
     title=#1,
     arc=2pt,
     boxrule=0.8pt,
     left=10pt, right=10pt, top=8pt, bottom=8pt,
     before skip=10pt, after skip=10pt
}

\newcommand{\algname}{PEARL}

\newcommand{\lift}[1]{{\color{red} \ensuremath{\blacktriangle} #1}}
\newcommand{\drop}[1]{{\color{blue} \ensuremath{\blacktriangledown} #1}}

\newcounter{promptcounter}
\renewcommand{\thepromptcounter}{\arabic{promptcounter}}
\newtcolorbox{MyBox}[2][]{%
  enhanced,
  breakable,
  colback=gray!5,
  colframe=gray!80!black,
  boxrule=1pt,
  toprule=2pt,
  rounded corners,
  arc=2pt,
  top=1.7mm,
  bottom=1.7mm,
  left=3mm,
  right=3mm,
  fuzzy shadow={0pt}{-2pt}{-0.5pt}{0.5pt}{black!35},
  title={\normalsize Prompt~\thepromptcounter.~#2}, 
  #1 
}

\title{PEARL: Front-Loading Relational Chains for Multi-Hop Table Retrieval}

\author{
Subeen Ho \quad Hyeongu Kang \quad SeongKu Kang \quad Susik Yoon  \\
Computer Science and Engineering \\
Korea University, Seoul, Korea \\
\texttt{\{hosubin02, hyeongukang, seongkukang, susik\}@korea.ac.kr}
}

\begin{document}

\clearpage

\maketitle

\begin{abstract}
While large language models (LLMs) have shown strong capabilities in tabular reasoning, retrieving relevant tables remains challenging due to the fragmented and relational structure of real-world data. Existing work typically relies on whole table representations that overlook cross-table semantics induced by join relationships. We propose PEARL, a training-free framework that shifts the paradigm toward vertical partitioning-based sub-table encoding. PEARL augments the retrieval corpus offline by generating multi-hop queries over pre-identified join paths and reorganizing relevant columns into vertically partitioned corpus units, enabling effective multi-table retrieval without query-time LLM inference. Experiments show that PEARL consistently outperforms existing methods, with up to +30.05\% gains in R@2 on 3-hop queries. The source code is available at \url{https://github.com/SOOB2NHO/PEARL}.
\end{abstract}

\section{Introduction}
Tables serve as one of the most effective and universal forms of knowledge in diverse domains~\cite{target}. Identifying tables relevant to a given query or task is a fundamental prerequisite for numerous downstream applications such as question answering~\cite{DTR}, fact verification~\cite{fact_verification}, and reasoning~\cite{chain-of-table, TableLLM}. Table retrieval has recently become more important with the rapid adoption of Retrieval-Augmented Generation (RAG)~\cite{RAG1, TableRAG} with a large language model (LLM) over structured knowledge bases, where the quality of retrieved tables directly affects the accuracy and reliability of generated responses.

While most existing studies on tabular learning simply assume that relevant tables are available at inference time~\cite{DTR, fact_verification, chain-of-table, TableLLM}, real-world tabular data is typically scattered across diverse organizational units with varying schemas and scales~\cite{tabular_importance}. In modern data lake environments~\cite{datalake1, datalake2, datalake3}, which contain a massive number of tables with heterogeneous and high-dimensional attributes, conventional retrieval approaches find difficulties in capturing fine-grained semantic relevance between queries and tables~\cite{jar}, making table retrieval a critical bottleneck for downstream applications.


\begin{figure}[bt!]
  \includegraphics[width=\columnwidth]{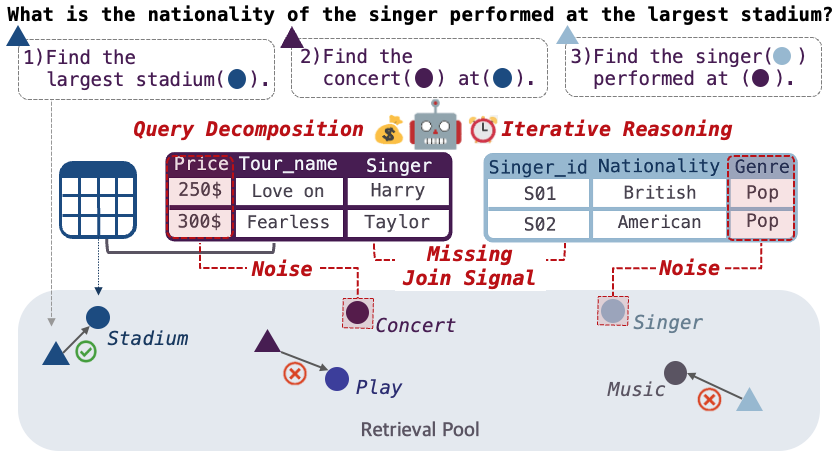}
  \caption{Limitations of existing approaches in multi-hop table retrieval from heterogeneous tables.}
  \label{fig:motivation}
\end{figure}

As exemplified in Figure~\ref{fig:motivation}, practical queries often require reasoning across multiple tables. Existing table retrieval approaches, however, suffer from fundamental limitations in handling such multi-table retrieval scenarios, as they rely on \textit{inference-time} query decomposition and iterative reasoning paired with offline \textit{whole-table} indexing~\cite{mmqa, jar, gjar}. In contrast to the tables that can be explicitly retrieved from a given query (e.g., \texttt{Stadium}), implicit (e.g., \texttt{Concert}) or deeply nested (e.g., \texttt{Singer}) ones within the expected multi-hop join path are challenging to identify. This is because query-irrelevant columns are intermingled within the whole-table embeddings, introducing non-trivial noise that hinders a model from capturing the latent join relationships implied by the query. 


Although decomposing the query or performing iterative reasoning at test time may alleviate the problem, such online workarounds introduce substantial computational cost as the number of tables and queries scales up, making them impractical in the real world. Moreover, these limitations intensify when explicit relational schemas are missing or incomplete~\cite{jar}, particularly within partial data lakes or raw table dumps.

In this work, we propose \textbf{\algname{}} (Path-awarE Augmented Retrieval over Linked schema), a simple yet effective multi-hop table retrieval framework that shifts the paradigm from online reasoning over whole-table indexing toward vertical partitioning-based pre-augmented indexing. The key idea of \algname{} is to \textit{front-load the recurring, open-ended online computational burden into a fixed, predictable offline indexing phase}. Specifically, by leveraging LLMs to synthesize and pre-index augmented sub-tables via vertical partitioning, \algname{} captures potential query semantics and latent cross-table relationships inherent in the original tables beforehand. 

In brief, PEARL performs offline indexing by constructing an inter-table graph to discover multi-hop join paths covering both implicit and explicit relations and generating LLM-driven schema-aware queries over each path. It further reorganizes query-relevant columns into vertically partitioned sub-table corpus units that better capture cross-table semantics. The resulting augmented corpus enables standard dense retrievers to outperform existing query decomposition-based methods without training or online LLM inference.

Our key contributions and findings are summarized as follows:
\vspace{-0.2cm}
 \begin{itemize}[leftmargin=10pt, noitemsep]
    \item To the best of our knowledge, this is the first work to incorporate join-path-aware query generation and vertical partitioning into offline indexing for multi-hop table retrieval, effectively front-loading reasoning of relational table chains.
    \item 
    Comprehensive evaluation on three multi-hop table retrieval benchmarks demonstrates that \algname{} outperforms the strongest baseline by 5.84\% on average on Recall@k, with a notable gain of up to 30.05\% for 3-hop queries.
    \item By restricting LLM inference to the offline indexing stage, \algname{} reduces token consumption by 4.1$\times$ on average compared to the strongest baseline, while achieving up to 79\% reduction in total latency. Notably, the indexing cost remains constant, being robust to the inference workload.
\end{itemize}
\begin{figure*}[hbt!]
  \centering
  \includegraphics[width=0.95\linewidth]{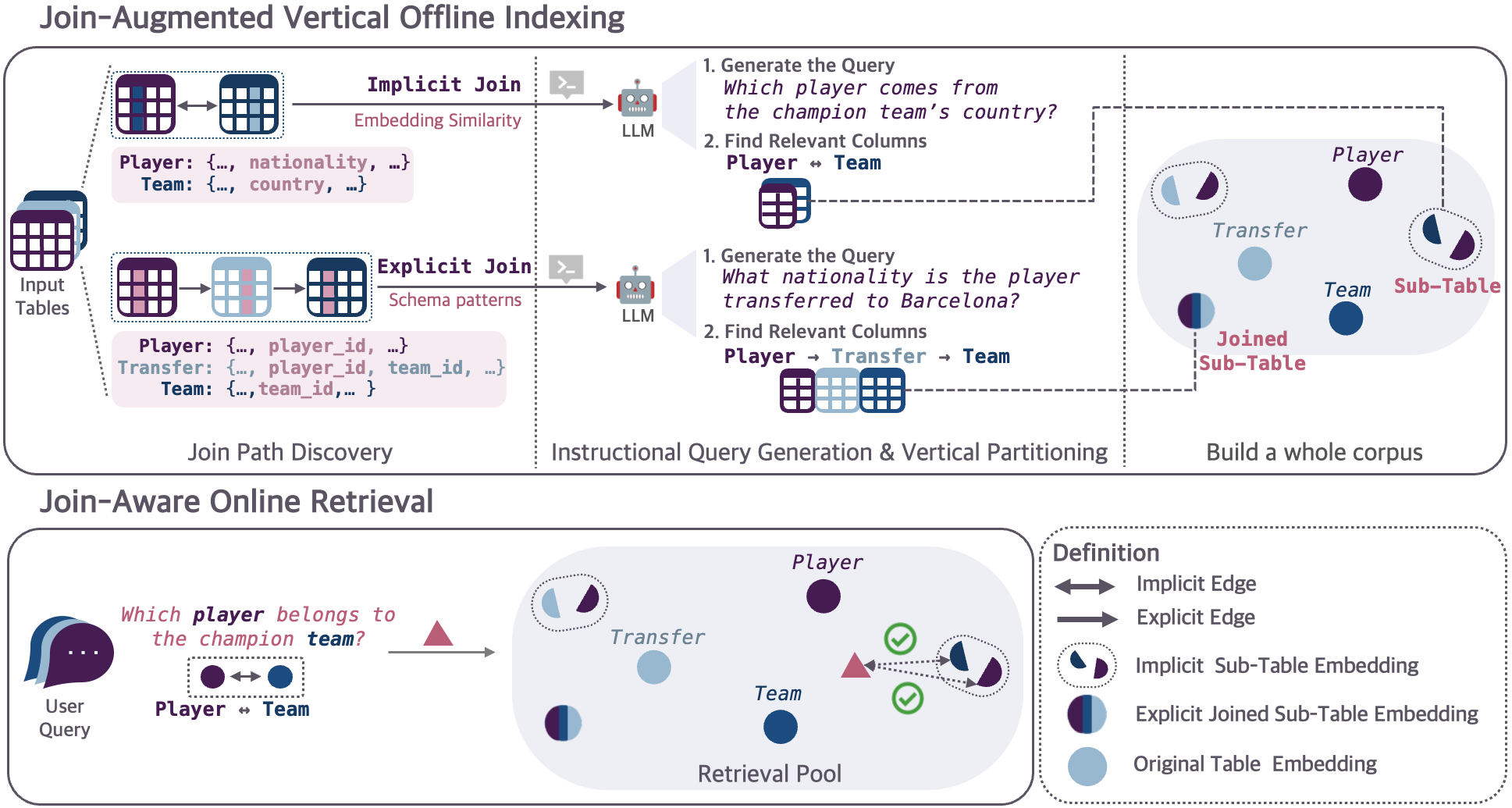} 
  \vspace{-0.2cm}
  \caption {The overall procedure of \algname{}.}
  \vspace{-0.3cm}
  \label{fig:framework}
\end{figure*}

\section{Preliminary}
The goal of multi-hop table retrieval is to identify the set of tables required to answer a query $q$, where the relevant tables may be connected via explicit or implicit join relationships inherent in the underlying schema structure. 

\paragraph{Explicit Join}
An explicit join captures a relationship that is structurally declared or strongly implied by the schema, such as shared column names, or high value containment between columns. These joins are grounded in schema-level constraints, representing the relational structure of the data.

\paragraph{Implicit Join.}
An implicit join captures a relationship that is latent in the data but not explicitly represented in the schema. For example, two columns in different tables may refer to the same real-world entity without any declared constraint (e.g., \texttt{customer} and \texttt{client}). Such joins can only be identified through value or semantic-level similarity of the table contents.

\section{Methodology}
\label{sec:methodology}

To efficiently model cross-table dependencies for multi-hop table retrieval, \algname{} identifies join paths and constructs vertically partitioned sub-tables for join-aware retrieval using a standard dense retriever, without additional training or online LLM inference. As illustrated in Figure~\ref{fig:framework}, \algname{} follows a two-stage pipeline consisting of Offline Indexing and Online Retrieval. 

During the offline indexing stage, \algname{} discovers join paths that capture how tables can be linked. For each selected path, we generate instructional queries using an LLM and apply vertical partitioning to retain only query-relevant information. This process transforms raw tables into structured corpus units that encode both individual tables and inter-table relationships. At the online retrieval stage, \algname{} encodes the user query and retrieves relevant tables by matching it against the join-aware augmented corpus.

\subsection{Join-Augmented Vertical Offline Indexing}
\label{sec:Offline}


\algname{} exploits LLM-based reasoning, confined within an offline stage in a constant and predictable manner. The offline indexing pipeline consists of three stages: Join Path Discovery, Instructional Query Generation, and Vertical Partitioning and Indexing, with a detailed procedure provided in Algorithm~\ref{alg:offline_indexing} in Appendix~\ref{app: Offline}.

\subsubsection{Join Path Discovery}
To enable join-aware query generation, we first construct join paths over the corpus $\mathcal{C}$, consisting of all tables in the dataset. Given a corpus $\mathcal{C}$, Join Path Discovery aims to identify joinable table pairs without schema-level annotations, reflecting real-world retrieval settings. The procedure consists of four phases: Index Construction, Candidate Discovery, Path Construction and Path Selection.

\paragraph{Index Construction}
\label{para:index_construction}
Before candidate search, we compute two indexes over all columns in $\mathcal{C}$: (1) a column embedding index and (2) a value embedding index. Both are used for similarity scoring in subsequent stages (Appendix~\ref{app:index_construction} provides details).


\paragraph{Candidate Discovery}
\label{para:candidate_discovery}
A candidate edge $e = (c_i, c_j)$ is a column pair drawn from two distinct tables $T_i$ and $T_j$ with a potential latent join relationship. We collect candidate edges via three complementary signals: structural patterns ($\mathcal{P}_S$), column similarity ($\mathcal{P}_B$), and cell value-overlap ($\mathcal{P}_V$), and define the final candidate edge pool as $\mathcal{P} = \mathcal{P}_S \cup \mathcal{P}_B \cup \mathcal{P}_V$. The formal definitions and detailed procedures for each signal are provided in Appendix~\ref{app:candidate_discovery}.


\paragraph{Path Construction}
Each candidate edge $e \in \mathcal{P}$ is categorized into two types. An explicit edge connects tables whose relationship is directly observable from schema-level signals, whereas an implicit edge captures semantically related tables without explicit structural connections. 

Paths are assigned to either $\mathcal{P}_{\mathrm{exp}}$ or $\mathcal{P}_{\mathrm{imp}}$ based on their edge composition, yielding two candidate pools passed to Path Selection. Specifically, paths composed entirely of explicit edges are assigned to $\mathcal{P}_{\mathrm{exp}}$, whereas those composed entirely of implicit edges are assigned to $\mathcal{P}_{\mathrm{imp}}$. Mixed paths are categorized based on whether the intermediate table serves as a structural bridge.


\paragraph{Path Selection}

We apply Lazy Greedy independently over $\mathcal{P}_{\mathrm{exp}}$ and $\mathcal{P}_{\mathrm{imp}}$ to select a diverse and high-quality set of join paths. Each candidate path $p$ is assigned a base score $f(p)$ by aggregating three complementary signals. Specifically, $S_{\mathrm{str}}$ captures structural anchors such as foreign-key patterns and identical column names, $S_{\mathrm{sem}}$ measures semantic similarity, and $S_{\mathrm{ovlp}}$ captures value overlap through Jaccard similarity and containment.
\begin{equation}
    f(p) = 
    \begin{cases} 
        S_{\mathrm{str}} + S_{\mathrm{ovlp}} & \text{if } p \in \mathcal{P}_{\mathrm{exp}}, \\ 
        S_{\mathrm{sem}} + S_{\mathrm{ovlp}} & \text{if } p \in \mathcal{P}_{\mathrm{imp}}.
    \end{cases}
\end{equation}

Starting from an empty set $\mathcal{P}^*$, the marginal gain $g(p \mid \mathcal{P}^*)$ for adding a path $p$ is defined as:
\begin{equation}
    g(p \mid \mathcal{P}^*) = f(p) + \lambda_{\mathrm{div}} \cdot \big| \mathcal{V}(p) \setminus \mathcal{V}(\mathcal{P}^*) \big|,
\end{equation}
where $\mathcal{V}(\mathcal{P}^*)$ denotes the tables covered by the previously selected paths, and $\mathcal{V}(p)$ denotes the tables covered by the candidate path $p$. We iteratively select the path $p^{'}$ that maximizes this marginal gain until the budget is exhausted:
\begin{equation}
    p^{'} = \underset{p \notin \mathcal{P}^*}{\arg\max} \; g(p \mid \mathcal{P}^*), \quad \text{for } |\mathcal{P}^*| < K.
\end{equation}
Both $\lambda_{\mathrm{div}}$ (controlling the diversity-quality balance) and $K$ (the corpus budget) are hyperparameters, which are further discussed in Section~\ref{sec: Sensitivity Analysis}.


\subsubsection{Instructional Query Generation}
\label{subsec: Query Generation}

To capture multi-hop information needs that span multiple tables, we generate discriminative queries over selected join paths $\mathcal{P}^*$. Given a join path $p \in \mathcal{P}^*$, we prompt an LLM with the full path structure and sampled rows from all tables in $p$. The LLM jointly generates a query that requires all tables in the path, together with query-relevant column subsets for each table, identifying the minimal columns necessary to answer the query.
\begin{equation}
    \bigl(q,\, \{c_i\}_{i=1}^{h}\bigr) = \mathrm{LLM}\!\left(p,\, \{R_i\}_{i=1}^{h},\, \mathcal{I}\right),
\end{equation}
where each $c_i \subseteq \mathcal{S}_i$ denotes the minimal set of columns from table $T_i$ necessary to answer $q$, $\mathcal{S}_i$ is the full schema of $T_i$, $R_i$ denotes sampled rows from $T_i$, and $h$ is the number of tables in the path. $\mathcal{I}$ denotes the task instruction. We impose a constraint that the query requires all tables in $p$; removing any $T_j \in p$ renders the query unanswerable. Task instructions for implicit and explicit paths are provided in Appendix~\ref{appendix:prompt}.


\subsubsection{Vertical Partitioning and Indexing}
\label{subsec: Vertical Partitoning}


Wide tables often contain many columns irrelevant to a given query, which dilutes the retrieval signal when the full schema is embedded as a single representation. To address this, \algname{} performs vertical partitioning to reorganize tables and generated queries into representations optimized for both structural and semantic matching. Given the query-relevant column subsets $c_i$ from the query generation stage, we perform vertical partitioning over the join path $p$, constructing query-conditioned sub-tables $\hat{T}_i$ by projecting each table $T_i \in p$ onto its corresponding column subset:
\begin{equation}
    \hat{T}_i = \Pi_{c_i}(T_i), \quad i = 1, \dots, h.
\end{equation}
For implicit paths, we construct sub-entries that capture pairwise semantic joins between tables. Each projected sub-table $\hat{T}_i$ is indexed as an independent retrieval unit and combined with its associated instructional queries, resulting in multiple embeddings per path. This enables multiple semantic views of each table within the path.
\begin{equation}
    \mathbf{e}_{\mathrm{imp},i} = \mathrm{Enc}\!\left(q \,\|\, \hat{T}_i\right), \quad i = 1, \dots, h.
\end{equation}
For explicit paths, we construct joined entries that capture structural joins between tables. We unify all projected sub-tables within a path together with the shared query, forming a single multi-table representation. This enables modeling of inter-table relations within a unified representation.
\begin{equation}
    \mathbf{e}_{\mathrm{exp}} = \mathrm{Enc}\!\left(q \,\|\, \hat{T}_1 \,\|\, \cdots \,\|\, \hat{T}_h\right).
\end{equation}
Both embedding types are stored in offline stage. This design narrows the semantic gap between the indexed table representations and actual user needs, while ensuring scalable retrieval via standard approximate nearest neighbor (ANN) search.


\subsection{Join-Aware Online Retrieval}
\label{sec:Online}

At online retrieval time, \algname{} uses only precomputed embeddings without additional LLM inference or reasoning. A query $q$ is scored against a unified retrieval pool $\mathcal{U}(T)$ consisting of implicit embeddings $\mathbf{e}_{\mathrm{imp},i}$, explicit embeddings $\mathbf{e}_{\mathrm{exp}}$, and original table embeddings $\mathbf{e}_{\mathrm{orig}}$ derived from join paths containing table $T$. The relevance score of a table $T$ is computed by max-pooling over all embeddings associated with $T$:
\begin{equation}
    \mathrm{Score}(q, T) = \max_{\mathbf{e} \in \mathcal{U}(T)}
    \cos\!\left(\mathrm{Enc}(q), \mathbf{e}\right).
\end{equation}
We retrieve the top-$k$ tables ranked by $\mathrm{Score}(q, T)$, enabling training-free retrieval without schema annotations and supporting efficient inference at query time.


\section{Experiments}\label{sec:experiments}
We comprehensively evaluate \algname{} on three multi-hop table retrieval benchmarks against established baselines to address seven research questions. The key takeaways are highlighted as follows:

\begin{itemize}[leftmargin=10pt, noitemsep]
    \item \textbf{Overall Performance Results (RQ1, RQ2, RQ3).}
    \algname{} consistently improves multi-hop retrieval accuracy, especially in higher-hop settings, while significantly reducing online inference cost via offline join-augmented indexing.

    \item \textbf{Ablation Study (RQ4, RQ5).}
    The gains of \algname{} are not driven by simple corpus expansion, but by join-augmented, vertical indexing that captures inter-table relationships. Moreover, explicit and implicit join representations provide complementary signals, and removing either consistently degrades performance.

    \item \textbf{Sensitivity Analysis (RQ6, RQ7).}
    Performance of \algname{} saturates as $K$ increases, with most gains achieved using only half of the join paths. $\lambda_{div}$ shows that higher diversity benefits complex multi-table queries. Overall, \algname{} is robust to hyperparameter variations, supporting practical deployment in real-world settings.
\end{itemize}

\subsection{Experimental setup}
\label{sec: setup}

\paragraph{Datasets and Evaluation Metrics}
\label{para:Dataset & Metrics}
We evaluate \algname{} on three multi-hop table retrieval benchmarks: BIRD~\cite{bird}, SPIDER~\cite{spider}, and MMQA~\cite{mmqa}. Since our method requires no training phase, we can apply it directly to this pool without any dataset-specific supervision. For MMQA, whose tables are not partitioned into separate databases, we use the entire corpus as a unified retrieval pool. For SPIDER and BIRD, we construct a unified corpus by removing database boundaries. We use the same dev split for evaluation as JAR and Greedy-JAR, but perform retrieval over the unified corpus rather than within individual databases. This setup removes the assumption that gold tables are confined to a known database, resulting in a more challenging and realistic search space that better reflects real-world data lakes, where tables from heterogeneous sources are pooled together. We report standard R@k for evaluating retrieval performance.

To discuss the potential impact of augmented tables in evaluating retrieval results, we also report adjusted recall, namely Slot R@$k$, which collapses all unit-level scores to the table level via the same max-pooling strategy. Additional details and evaluation results are provided in Appendix~\ref{app: adjusted-recall}.



\paragraph{Baselines and Implementation Details}
We compare \algname{} with baselines across three paradigms:
(1) \emph{text models}, including stella-v5~\cite{stella}, which demonstrates strong table-to-text alignment on the TARGET benchmark~\cite{target}, and contriever~\cite{contriever}, the backbone of decomposition-based retrieval models;
(2) \emph{Single-hop retrieval model}, including DTR~\cite{DTR}, a strong single-hop table retrieval model; and (3) \emph{multi-hop table retrieval models}, including JAR~\cite{jar},
Greedy-JAR~\cite{gjar}, CORE-T~\cite{CORE-T}, and MURRE~\cite{murre},
which are available with reproducible implementations. All baselines are reproduced using the authors’ publicly released implementations. 

For architectural consistency, \algname{} adopts Stella-v5 as its backbone encoder and Llama 3.3 70B~\cite{llama} as the query generation LLM, with temperature $0.7$, a maximum of $1{,}024$ output tokens, and ten sampled rows per table as input context for query generation. The diversity-quality balance weight $\lambda_{\mathrm{div}}$ is fixed at 0.1, and the corpus budget $K$ is set to $\lfloor|\mathcal{C}| / 2 \rfloor$ across all datasets. All methods are evaluated under a uniform 512-token input-length constraint.

\begin{table*}[t]
\centering
\scriptsize 
\setlength{\tabcolsep}{6pt}
\renewcommand{\arraystretch}{0.95}
\resizebox{\textwidth}{!}{
\begin{tabular}{ll cc|cc|cc}
\toprule
\multirow{2}{*}{\textbf{Setting}} & \multirow{2}{*}{\textbf{Model}} & \multicolumn{2}{c|}{\textbf{SPIDER}} & \multicolumn{2}{c|}{\textbf{MMQA}} & \multicolumn{2}{c}{\textbf{BIRD}} \\ 
\cmidrule(lr){3-4} \cmidrule(lr){5-6} \cmidrule(lr){7-8}
 & & R@2 & R@5 & R@2 & R@5 & R@2 & R@5 \\ \midrule
\multirow{9}{*}{\textbf{3-Hop}} 
 & Contriever          & 56.10 & 91.10 & 38.10 & 55.50 & 45.80 & 78.70 \\
 & Stella-v5           & 58.89 & 87.78 & \underline{47.30} & \underline{71.70} & 52.17 & 85.33 \\ 
 & JAR                 & \underline{62.80} & \underline{94.40} & 44.70 & 64.20 & 54.30 & 86.20 \\
 & Greedy-JAR          & 60.00 & 90.60 & 44.50 & 61.80 & 54.30 & \textbf{88.30} \\
 & DTR                 & 56.67 & 89.44 & 44.39 & 66.46 & 48.33 & 80.83 \\
 & MURRE               & 52.78 & 85.00 & 34.05 & 56.69 & 45.83 & 74.83 \\
 & CORE-T              & 55.56 & 92.22 & 45.80 & 64.38 & 52.67 & 84.33 \\
 \cmidrule(lr){2-8}
 & \algname{}          & \textbf{81.67} & \textbf{100.00} & \textbf{58.09} & \textbf{80.43} & \textbf{59.17} & \underline{87.33} \\[0.5ex] 
 &                     & (\lift{30.05}\%) & (\lift{5.93}\%) & (\lift{22.81}\%) & (\lift{12.18}\%) & (\lift{8.97}\%) & (\drop{1.10}\%) \\ 
\midrule
\multirow{9}{*}{\textbf{2-Hop}} 
 & Contriever          & 78.20 & 97.80 & 49.10 & 63.60 & 62.00 & 84.00 \\
 & Stella-v5           & 76.97 & 96.06 & 55.76 & \underline{72.63} & 64.67 & 92.29 \\ 
 & JAR                 & 85.00 & \underline{98.10} & 55.90 & 65.20 & \textbf{80.30} & 92.50 \\
 & Greedy-JAR          & \underline{86.50} & 95.70 & \underline{56.80} & 64.00 & \underline{79.10} & \underline{93.70} \\
 & DTR                 & 80.28 & 96.18 & 52.47 & 66.37 & 62.62 & 87.38 \\
 & MURRE               & 74.80 & 92.22 & 49.15 & 70.20 & 56.20 & 82.46 \\
 & CORE-T              & 80.21 & 96.97 & 54.51 & 67.23 & 72.78 & 92.57 \\
 \cmidrule(lr){2-8}
 & \algname{}          & \textbf{89.57} & \textbf{99.49} & \textbf{60.57} & \textbf{75.58} & 72.81 & \textbf{94.38} \\[0.5ex] 
 &                     & (\lift{3.55}\%) & (\lift{1.42}\%) & (\lift{6.64}\%) & (\lift{4.06}\%) & (\drop{9.33}\%) & (\lift{0.73}\%) \\ 
\midrule
\multirow{9}{*}{\textbf{Unified}} 
 & Contriever          & 84.80 & 97.90 & 46.30 & 61.50 & 62.80 & 84.60 \\
 & Stella-v5           & 85.46 & 97.42 & 53.10 & 71.72 & 66.71 & 91.36 \\ 
 & JAR                 & 87.80 & \underline{98.20} & 53.10 & 64.60 & \textbf{75.10} & 90.80 \\
 & Greedy-JAR          & \underline{88.20} & 97.00 & \underline{53.80} & 63.20 & 74.40 & 91.70 \\
 & DTR                 & 86.99 & 97.45 & 50.36 & 65.96 & 65.80 & 87.78 \\
 & MURRE               & 77.27 & 92.60 & 45.62 & 66.82 & 58.84 & 82.51 \\
 & CORE-T              & 87.58 & 98.22 & 52.16 & 66.13 & \underline{74.71} & \underline{92.54} \\
 \cmidrule(lr){2-8}
 & \algname{}          & \textbf{91.25} & \textbf{99.33} & \textbf{58.84} & \textbf{76.21} & 73.64 & \textbf{93.36} \\[0.5ex] 
 &                     & (\lift{3.46}\%) & (\lift{1.13}\%) & (\lift{9.37}\%) & (\lift{6.26}\%) & (\drop{1.95}\%) & (\lift{0.89}\%) \\ 
\midrule
\multicolumn{8}{l}{\textit{Ablation of \algname{} (Unified)}} \\
\multicolumn{2}{l}{\hspace{3mm} w/o joined entry}   & 87.55 & 99.12 & 54.21 & 73.34 & 69.66 & 92.73 \\
\multicolumn{2}{l}{\hspace{7mm}{\tiny (= w/ implicit-only)}} & & & & & & \\
\multicolumn{2}{l}{\hspace{3mm} w/ explicit-only}   & 91.62 & 99.18 & 57.62 & 74.05 & 74.10 & 93.81 \\
\multicolumn{2}{l}{\hspace{3mm} Random-view}        & 84.66 & 97.31 & 52.39 & 70.80 & 66.25 & 90.78 \\
\multicolumn{2}{l}{\hspace{3mm} Random-query}       & 85.99 & 97.05 & 54.97 & 73.15 & 68.73 & 92.51 \\
\bottomrule
\end{tabular}%
}
\caption{Standard R@$k$ across 3-Hop, 2-Hop, and Unified configurations. Best results are in \textbf{bold}, second best \underline{underlined}. $\color{red}\blacktriangle$ and $\color{blue}\blacktriangledown$ indicate improvement and degradation over the strongest baseline.}
\label{tab:main_results}
\end{table*}
\begin{table}[t]
\centering
\scriptsize
\setlength{\tabcolsep}{3pt}
\renewcommand{\arraystretch}{0.95}
\resizebox{\columnwidth}{!}{
\begin{tabular}{l cc cc cc}
\toprule
\multirow{2}{*}{\textbf{Model}} & \multicolumn{2}{c}{\textbf{SPIDER}} & \multicolumn{2}{c}{\textbf{MMQA}} & \multicolumn{2}{c}{\textbf{BIRD}} \\
\cmidrule(lr){2-3}\cmidrule(lr){4-5}\cmidrule(lr){6-7}
 & R@2 & R@5 & R@2 & R@5 & R@2 & R@5 \\
\midrule
Stella-v5  & 85.46 & 97.42 & 53.10 & 71.72 & 66.71 & 91.36 \\
\algname{} & \textbf{88.48} & \textbf{98.75} & \textbf{54.26} & \textbf{72.88} & \textbf{69.66} & \textbf{92.73} \\
\bottomrule
\end{tabular}
}
\caption{Adjusted recall (Slot R@$k$) under the Unified setting. Full per-hop results are in Appendix~\ref{app: adjusted-recall}.}
\label{tab:slot_summary}
\vspace{-0.2cm}
\end{table}
\begin{table*}[t]
\vspace{-0.4cm}
\centering
\footnotesize 
\setlength{\tabcolsep}{0pt}
\begin{tabular*}{\textwidth}
{@{\extracolsep{\fill}} ll ccccc lccc}
\toprule
 & & \multicolumn{5}{c}{\textbf{Latency (s)}} & \multicolumn{4}{c}{\textbf{Token Consumption (K)}} \\
\cmidrule(lr){3-7} \cmidrule(lr){8-11}
\textbf{Data} & \textbf{Phase} & \textbf{0.25|C|} & \textbf{0.50|C|} & \textbf{0.75|C|} & \textbf{1.00|C|} & \textbf{GJAR} & \textbf{Type} & \textbf{\algname{}} & \textbf{GJAR} & \textbf{Ratio} \\
\midrule
\multirow{3}{*}{SPIDER}
 & Offline & 222 & 282 & 335 & 399 & —     & Prompt & 97  & 450 & 4.6$\times$ \\
 & Online  & 4   & 4   & 4   & 5   & 894   & Comp.  & 11  & 36  & 3.2$\times$ \\
 & \textbf{Total} & \textbf{226} & \textbf{286} & \textbf{339} & \textbf{404} & \textbf{894} & \textbf{Total} & \textbf{108} & \textbf{486} & \textbf{4.5$\times$} \\
\midrule
\multirow{3}{*}{MMQA}
 & Offline & 774 & 1,200 & 1,637 & 2,063 & —     & Prompt & 675 & 1,458 & 2.2$\times$ \\
 & Online  & 25  & 32    & 39    & 46    & 3,225 & Comp.  & 94  & 139  & 1.5$\times$ \\
 & \textbf{Total} & \textbf{799} & \textbf{1,232} & \textbf{1,675} & \textbf{2,109} & \textbf{3,225} & \textbf{Total} & \textbf{768} & \textbf{1,598} & \textbf{2.1$\times$} \\
\midrule
\multirow{3}{*}{BIRD}
 & Offline & 228 & 297 & 379 & 458 & —     & Prompt & 118 & 674 & 5.7$\times$ \\
 & Online  & 5   & 5   & 6   & 6   & 1,399 & Comp.  & 10  & 61  & 5.9$\times$ \\
 & \textbf{Total} & \textbf{233} & \textbf{302} & \textbf{385} & \textbf{464} & \textbf{1,399} & \textbf{Total} & \textbf{129} & \textbf{735} & \textbf{5.7$\times$} \\
\bottomrule
\end{tabular*}
\caption{Efficiency comparison of \algname{} and GJAR. Left: Latency breakdown (seconds). Right: Token consumption (K) for offline LLM calls. All efficiency measurements use GPT-4o mini as the LLM backbone for both evaluations. GJAR has no offline phase.}
\vspace{-0.1cm}
\label{tab:efficiency_combined}
\vspace{-0.4cm}
\end{table*}

\subsection{Overall Performance Results}
\label{sec: Overall results}

\textbf{RQ1. Does join-augmented vertical indexing improve multi-hop retrieval accuracy?}
Table~\ref{tab:main_results} summarizes retrieval performance across SPIDER, MMQA, and BIRD, where \algname{} achieves the best performance in most cases, outperforming the strongest baseline by 3.46\% on SPIDER and 9.37\% on MMQA under the unified setting. On BIRD, we observe minor fluctuations across configurations, which we analyze in Appendix~\ref{app: BIRD}.

Since \algname{} is built on the Stella-v5 encoder, the performance gap between the two isolates the effect of vertical partitioning. Under the unified setting, \algname{} improves over Stella-v5 by 6.77\% on SPIDER, 10.81\% on MMQA, and 10.39\% on BIRD in R@2. Notably, even the w/o joined entry variant, which independently embeds each sub-table unit without constructing joined entries, consistently outperforms Stella-v5 across all benchmarks. These results suggest that \algname{} reduces semantic noise by filtering irrelevant columns, improving retrieval performance. 

To further verify that these gains are not simply an artifact of joined entries covering multiple gold tables, we additionally report adjusted recall in Table~\ref{tab:slot_summary}. Under this unit-level metric, \algname{} consistently outperforms the Stella-v5 backbone across all three benchmarks in the unified setting, indicating that the improvements stem from more effective noise reduction rather than artificially inflated recall due to bundled gold tables. Full per-hop results are provided in Appendix~\ref{app: adjusted-recall}.

\textbf{RQ2. Does the effectiveness of \algname{} increase with query hop complexity?}


Table~\ref{tab:main_results} presents retrieval performance across 2-hop and 3-hop configurations. The performance gap between \algname{} and baseline methods consistently increases with the number of hops. In the 2-hop setting, \algname{} shows consistent improvements on SPIDER and MMQA, while a slight degradation is observed on BIRD. 

In the 3-hop setting, performance gains become more pronounced, with \algname{} outperforming the strongest baseline in R@2 by 30.05\%, 22.81\%, and 8.97\% on SPIDER, MMQA, and BIRD, respectively, indicating stronger benefits under higher cross-table reasoning complexity where join-aware vertical indexing better captures inter-table relational structure.

This trend is particularly pronounced for MURRE, an online iterative baseline. \algname{}'s margin over MURRE widens
with hop count across all three benchmarks. This suggests that precomputing join-aware, vertically partitioned units becomes increasingly effective as relational reasoning deepens, whereas the cost of online reasoning compounds with each additional hop.


\textbf{RQ3. Does \algname{} reduce overall latency and LLM cost compared to baselines?}

We investigate whether \algname{} reduces online LLM dependency compared to decomposition-based retrieval methods. Since Greedy-JAR consistently outperforms JAR in efficiency~\cite{gjar}, we use it as the representative baseline for latency comparison. 

\begin{figure}[t]
\centering
\includegraphics[width=\columnwidth]{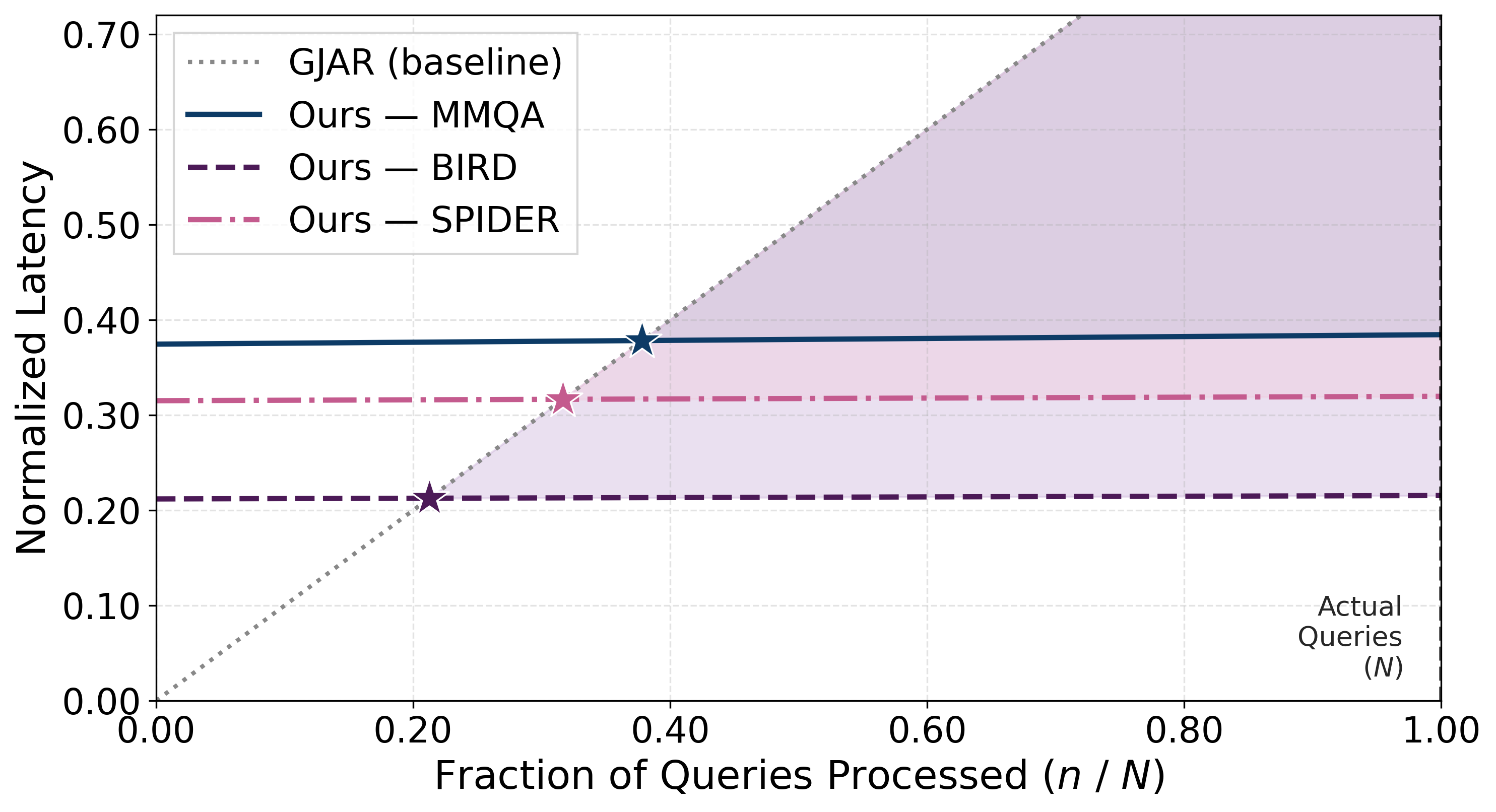}
\vspace{-0.4cm}
\caption{Latency (normalized by GJAR) vs.\ query fraction. Stars denote break-even points.}
\vspace{-0.4cm}
\label{fig:efficiency}
\end{figure}

As shown in Table~\ref{tab:efficiency_combined}, \algname{} shifts all LLM latency to the offline stage, reducing online latency by over 99\% across all datasets. Figure~\ref{fig:efficiency} shows that this design reaches a break-even point after processing only 22--35\% of the incoming query workload and achieves 62--79\% total latency reduction at full query volume, with efficiency gains from offline amortization becoming more pronounced as query volume increases. The storage and indexing overhead introduced by this offline design
remains modest and is analyzed in Appendix~\ref{app: overhead}.

We additionally measure token consumption to evaluate LLM usage efficiency. As shown in Table~\ref{tab:efficiency_combined}, \algname{} consistently achieves higher retrieval performance (Appendix~\ref{app:unified_llm}), while \algname{} requires fewer LLM tokens. Averaged across the three benchmarks, \algname{} reduces the total token consumption by 4.1$\times$ compared to Greedy-JAR. These results indicate that \algname{} is not only faster at query time but also more token-efficient than decomposition-based retrieval approaches. 


\subsection{Ablation Study}
\label{sec: Ablation study}
We conduct ablation studies to analyze the effect of join-aware augmentation and the roles of explicit and implicit representations in \algname{}.


\textbf{RQ4. Are the gains from PEARL affected by corpus expansion?}
\label{RQ4}
We compare \algname{} against two corpus-expansion variants. Random-view augments the corpus with randomly projected sub-tables as a corpus-expansion-only variant (Appendix~\ref{app:random_view}). Random-query augments each table with five discriminative questions, offering diverse semantic views without modeling join-aware relationships (Appendix~\ref{app: quocca}).

Despite producing a corpus of comparable size, Random-view performs worse than Stella-v5 across all datasets in Table~\ref{tab:main_results}. In contrast, \algname{} consistently outperforms both Stella-v5 and Random-view, surpassing Random-view on R@2 by 7.78\%, 12.31\%, and 11.15\% on SPIDER, MMQA, and BIRD, respectively. These results show that retrieval gains do not arise merely from increasing corpus size, but from join-aware augmentation enabled by vertical partitioning, which isolates and organizes join-relevant information.

Although Random-query expands the corpus more aggressively than \algname{}, \algname{} consistently outperforms it on R@2 across all datasets, achieving relative gains of 6.12\%, 7.04\%, and 7.14\% on SPIDER, MMQA, and BIRD, respectively. While Random-query improves the semantic coverage of individual tables, it lacks structural signals across tables. In contrast, \algname{} generates queries grounded in discovered join paths, modeling inter-table relationships. The same observation holds when compared with CORE-T, an offline augmentation method that attaches LLM-generated context to whole-table representations. \algname{} outperforms it on five of the six unified metrics, most notably by 12.81\% on R@2 and 15.24\% on R@5 for MMQA. These results suggest that structural awareness is more important than increasing semantic diversity alone in multi-hop scenarios.


\begin{figure}[t]
\centering
\includegraphics[width=\columnwidth]{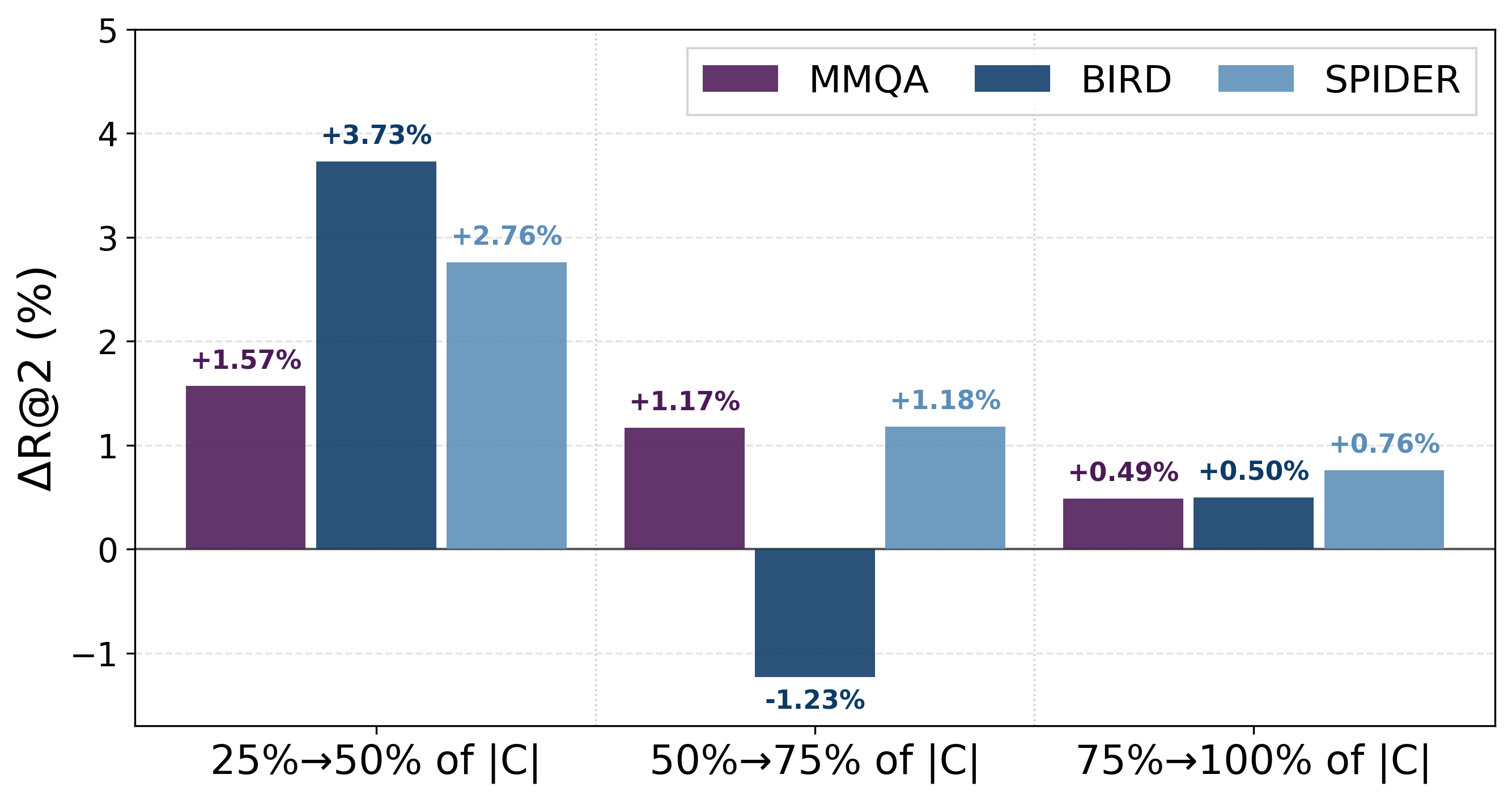}
\vspace{-0.2cm}
\caption{Marginal R@2 gain ($\Delta$R@2) per corpus-budget increment under the full augmentation setting across three datasets.}
\vspace{-0.2cm}
\label{fig:corpus_gain}
\end{figure}


\textbf{RQ5. Are both explicit and implicit join necessary for effective multi-hop retrieval?}

We compare \algname{} with variants that retain only explicit or implicit join representations. As shown in Table~\ref{tab:main_results}, removing either component consistently degrades performance. Explicit-only variants outperform implicit-only ones on all three datasets (SPIDER +4.65\%, MMQA +6.29\%, BIRD +6.37\%), indicating that joined entries provide a stronger individual signal than sub-entries alone. Nevertheless, neither variant consistently matches the full \algname{} across datasets under Slot R@2 (Table~\ref{tab:slot}), confirming that joined entries anchor structural chains, while sub-entries surface latent semantic links, making both signals essential for robust multi-hop retrieval. Appendix~\ref{app: asymmetric} further validates the asymmetric encoding, by encoding implicit paths as joined entries.


\subsection{Sensitivity Analysis}
\label{sec:  Sensitivity Analysis}
We analyze the sensitivity of \algname{} to two key hyperparameters: the corpus budget $K$ and the diversity-quality balance weight $\lambda_{\mathrm{div}}$.


\textbf{RQ6. How does retrieval performance vary with the corpus budget $K$?}
Figure~\ref{fig:corpus_gain} reports the marginal R@2 gain as the fraction of sampled join paths increases from 25\% to 100\% of the corpus size $|\mathcal{C}|$. The largest gains are observed when sampling the first half of join paths (25\% to 50\%), with improvements of 1.57\%, 3.73\%, and 2.76\% on MMQA, BIRD, and SPIDER, respectively. Beyond this point, the marginal benefit decreases substantially and approaches saturation in the 75\% to 100\% range, with BIRD even showing a slight performance drop between 50\% and 75\%. These results suggest that sampling half of the available join paths is sufficient to capture most retrieval gains, justifying it as the default $K$ in all experiments.

\textbf{RQ7. How does the diversity-quality balance weight $\lambda_{\mathrm{div}}$ affect retrieval performance?}

\begin{figure}[t]
\vspace{-0.2cm}
\centering
\includegraphics[width=\columnwidth]{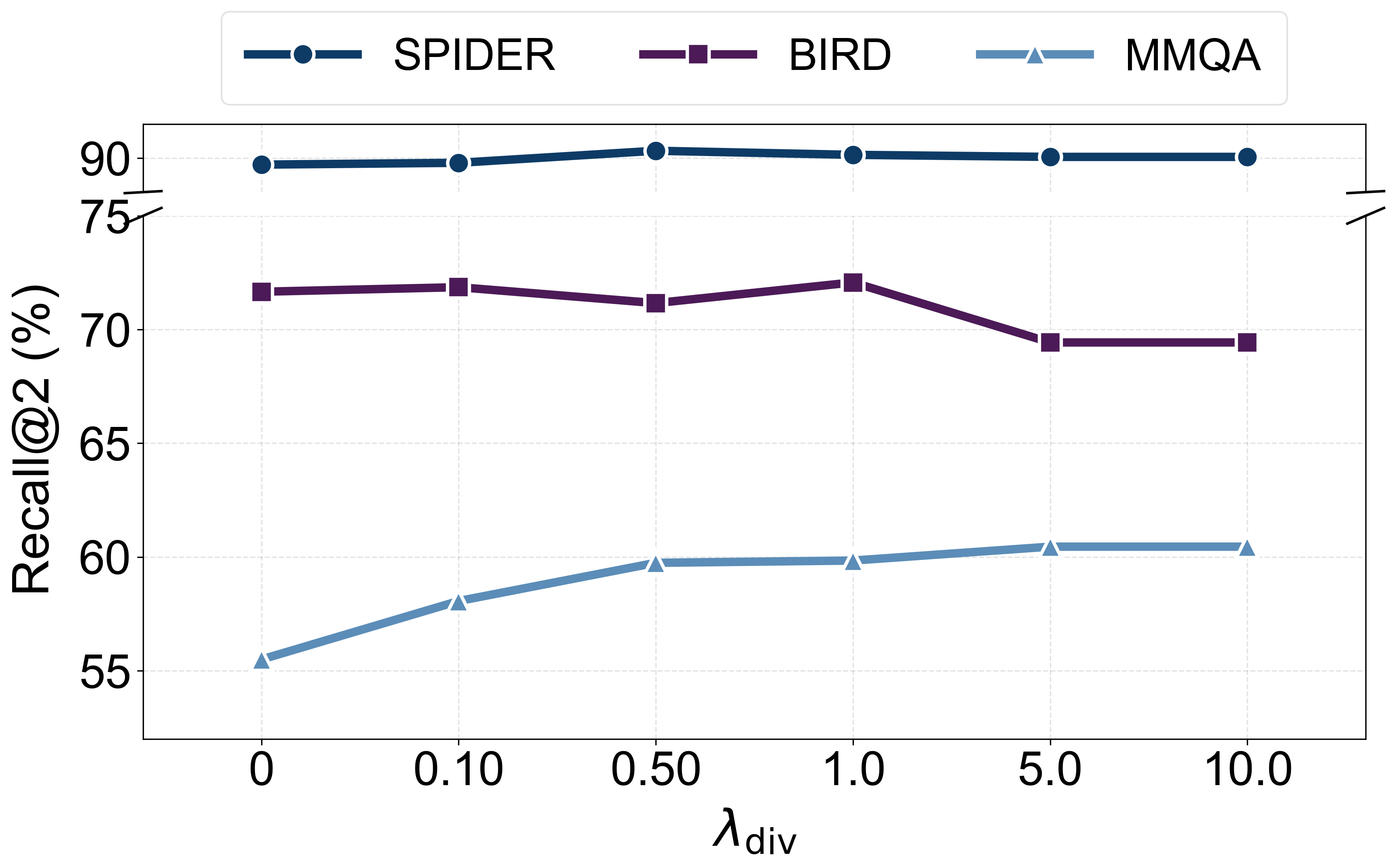}
\vspace{-0.6cm}
\caption{Sensitivity analysis on the diversity-quality trade-off hyperparameter $\lambda_{\mathrm{div}}$ across three benchmarks.}
\vspace{-0.5cm}
\label{fig:lambda}
\end{figure}

Figure~\ref{fig:lambda} reports performance across 
varying $\lambda_{\mathrm{div}}$. Since $\lambda_{\mathrm{div}}$ controls the balance between query diversity and join quality, the sensitivity patterns reflect the relative importance of each factor per dataset. MMQA consistently improves as $\lambda_{\mathrm{div}}$ increases, suggesting that broader query coverage is particularly important in its complex multi-table retrieval setting. In contrast, BIRD shows sharp performance drop at higher values, indicating that join quality is more dominant and excessive diversity is detrimental. SPIDER remains relatively insensitive across different values, suggesting that its retrieval performance is stable across different diversity-quality balances. Importantly, all three datasets  converge to identical performance beyond $\lambda_{\mathrm{div}} = 5.0$, indicating that excessively strong diversity penalization provides no additional benefit.

\section{Related Work}
\label{sec:related_work}


With the rapid proliferation of large-scale data lakes integrating heterogeneous sources, table retrieval has become a critical problem for identifying relevant tables given a query~\cite{datalake1, datalake2, datalake3}. Beyond single-table retrieval, recent work has explored multi-hop table retrieval, which requires jointly retrieving multiple related tables. We distinguish two types of join relationships, explicit and implicit, and categorize existing methods based on how they model inter-table relationships.


\paragraph{Explicit Join Modeling}
Explicit join-based approaches leverage schema information to identify related table sets or discover candidate joins across tables~\cite{jar, gjar, CORE-T, birdie}. These methods exploit explicit relational structure to narrow the search space before downstream reasoning.

Representative methods explore join candidates using metadata associated with natural language queries~\cite{birdie} or structurally model foreign-key relationships and schema consistency across tables~\cite{jar, gjar, CORE-T}. However, these approaches often rely on monolithic whole-table encoding before passing tables to the model, which often exceeds the context window limitations of LLMs in real-world data lake environments with massive numbers of columns and rows. Moreover, indiscriminately incorporating irrelevant columns introduces substantial semantic noise into the retrieval process.
\vspace{-0.1cm}
\paragraph{Implicit Join Alignment}
Considering real-world environments where structural information is incomplete or not explicitly available~\cite{chain-of-table}, prior work has increasingly explored implicit semantic alignment approaches that leverage semantic similarities across data or the reasoning capabilities of LLMs, such as dynamically exploring retrieval paths at query time~\cite{mmqa, FGTR}. Unlike explicit join modeling that relies on predefined relational metadata, these methods attempt to infer latent inter-table relationships through iterative reasoning and semantic matching. However, they typically require repeated LLM invocations for each query, resulting in substantial latency and computational cost as the number of tables and query complexity increase.

Related topics beyond the scope of this section, including single-hop table retrieval and LLM-based tabular reasoning, are discussed in Appendix~\ref{app:additional_related_work}.
\vspace{-0.1cm}
\section{Conclusion}
\vspace{-0.15cm}
We presented \algname{}, a table retrieval framework that bridges the semantic gap between tables and multi-hop queries by constructing a join-aware corpus. By generating implicit sub-table views and explicit joined entries for structurally or semantically related table pairs, \algname{} enables a standard embedding model to retrieve multi-hop evidence without any online LLM involvement or iterative query decomposition.

Experiments on SPIDER, BIRD, and MMQA demonstrate that \algname{} consistently improves retrieval performance, beyond simple corpus augmentation. Furthermore, \algname{} reduces online latency by over 99\% and total latency by up to 79\%, while cutting token consumption by up to 4.1$\times$ on average compared to Greedy-JAR, and its efficiency advantage grows as query volume scales.

\section*{Limitations}

While \algname{} demonstrates strong retrieval performance, table retrieval remains an open problem for complex queries that require reasoning over multiple joinable tables. We currently limit join paths up to a maximum length of three, which provides sufficient coverage of most benchmark cases. However, real-world settings may involve longer dependency chains. We leave the exploration of more scalable path construction and richer inter-table connections as future work.


As noted in Appendix~\ref{app: BIRD}, although \algname{} is fully training-free and can be applied directly to an entire table corpus without dataset-specific supervision, our evaluation is limited to academic benchmarks. Further case studies on real-world domain-specific scenarios would be a promising direction to explore the practical applicability and robustness in diverse deployment settings.

From a deployment perspective, \algname{} maintains a larger index than whole-table retrieval, as each selected join path introduces additional corpus units. Although this overhead is bounded by the corpus budget K and remains modest in our experiments (Appendix~\ref{app: overhead}), memory-constrained deployments may require a smaller budget. Exploring the trade-off between storage efficiency and retrieval effectiveness is an interesting direction for future work. As our incremental indexing evaluation with a single corpus expansion step in Appendix~\ref{app: dynamic} suggests that \algname{} can efficiently update the index without full re-indexing, its behavior under continuous corpus evolution is worth further exploration. Addressing this challenge is closely related to developing more flexible offline indexing strategies. More broadly, \algname{} adopts a fully offline indexing pipeline, precomputing all selected paths before deployment. A hybrid workflow that precomputes only high-value paths while resolving the remaining paths at query time could provide a better balance between indexing cost and query efficiency, as discussed in Appendix~\ref{app: adaptive}.



\smallskip
\section*{Acknowledgments}
This work was supported by Korea University--KT (Korea Telecom) R\&D Center, the Institute of Information \& Communications Technology Planning \& Evaluation (IITP) grants (IITP-2026-RS-2020-II201819, IITP-2026-RS-2024-00436857, IITP-2026-RS-2025-02304828) and the National Research Foundation of Korea (RS-2024-00406320).

\newpage
\bibliography{References}

\newpage
\appendix
\section{Additional Related Work}

\label{app:additional_related_work}

\subsection{Single-hop Table Retrieval}
\label{app:singlehop}

Unlike real-world user queries, which often require reasoning over relationships across multiple tables~\cite{userquery}, most existing studies have primarily focused on single-table retrieval scenarios. Traditional keyword-based approaches~\cite{SPLADE, BM25} and embedding-based retrieval models~\cite{TAPAS, RIM, DERT, DTR} typically adopt a single table representation strategy, where the entire schema and sampled rows are compressed into a single vector representation, limiting their ability to capture inter-table relationships. Consequently, they struggle to effectively model the multi-table reasoning requirements commonly observed in real-world open-domain environments~\cite{jar}.

\subsection{LLM Reasoning on Tabular Data}
\label{app:llm_reasoning}

Recent studies have explored approaches where LLMs directly transform retrieved tables~\cite{chain-of-table} or decompose tables into finer-grained units for more precise matching and reasoning~\cite{dater, FGTR}. Although these methods achieve strong reasoning performance across various tabular tasks, they still suffer from substantial computational overhead due to repeated online LLM inference for each user query, limiting their applicability for latency-sensitive applications.

To mitigate this, prior work has proposed leveraging LLMs during the offline stage to generate synthetic queries for retrieval augmentation~\cite{cgpt}. However, such approaches still fundamentally fail to resolve the semantic noise introduced by irrelevant columns and remain confined to single-table retrieval settings, making effective modeling of inter-table relationships difficult.


\section{Details of Offline Corpus Augmentation}
\label{app: Offline}

This section describes the implementation details of \algname{}'s offline corpus augmentation pipeline, summarized in Algorithm~\ref{alg:offline_indexing}.



\subsection{Join Path Discovery: Implementation Details}
\label{appendix:find_join}

Join path discovery consists of four stages: index construction, candidate discovery, path construction, and path selection. Each stage is described in detail in the following subsections.


\subsubsection{Details of Index Construction}
\label{app:index_construction}

\begin{itemize}

    \item \textbf{Column Embedding Index:} Normalized column names are indexed with FAISS for approximate nearest-neighbor search (ANN), supporting Candidate Discovery ($\mathcal{P}_B$) and semantic scoring $S_{\mathrm{sem}}$ in Path Selection.


    \item \textbf{Value Embedding Index:} Five sampled cell values per column are encoded and mean-pooled into a single value embedding for semantic scoring $S_{\mathrm{sem}}$ in Path Selection. 
\end{itemize}


\begin{algorithm}[t]
\DontPrintSemicolon
\SetAlgoLined
\footnotesize
\caption{Offline indexing of \algname{}}
\label{alg:offline_indexing}

\KwIn{Table corpus $\mathcal{C}$, LLM instructions $\mathcal{I}_{\mathrm{imp}}, \mathcal{I}_{\mathrm{exp}}$, budget $K$, diversity weight $\lambda_{\mathrm{div}}$}
\KwOut{Augmented corpus $\mathcal{U}$}

\BlankLine
\textbf{Initialize} $\mathcal{U} \leftarrow \emptyset$ \;

\BlankLine
\tcp{Step 1: Join Path Discovery}
$\mathcal{P} \leftarrow \mathcal{P}_S \cup \mathcal{P}_B \cup \mathcal{P}_V$ \;
\tcp{Candidate Discovery}
Classify $\mathcal{P}$ into $\mathcal{P}_{\mathrm{exp}},\, \mathcal{P}_{\mathrm{imp}}$ \;
\tcp{Path Construction}
$\mathcal{P}^* \leftarrow \mathrm{LazyGreedy}(\mathcal{P}_{\mathrm{exp}}, \mathcal{P}_{\mathrm{imp}}; K, \lambda_{\mathrm{div}}, h \in \{2,3\})$ \;
\tcp{Path Selection}

\BlankLine
\For{each path $p = (T_1, \dots, T_h) \in \mathcal{P}^*$}{

    \tcp{Step 2: Instructional Query Generation}
    \eIf{$p \in \mathcal{P}_{\mathrm{imp}}$}{
        $\bigl(q,\, \{c_i\}_{i=1}^{h}\bigr) \leftarrow \mathrm{LLM}(p,\, \{R_i\}_{i=1}^{h},\, \mathcal{I}_{\mathrm{imp}})$ \;
    }{
        $\bigl(q,\, \{c_i\}_{i=1}^{h}\bigr) \leftarrow \mathrm{LLM}(p,\, \{R_i\}_{i=1}^{h},\, \mathcal{I}_{\mathrm{exp}})$ \;
    }

    \BlankLine
    \tcp{Step 3: Vertical Partitioning and indexing}
    $\hat{T}_i \leftarrow \Pi_{c_i}(T_i)$ \textbf{for each} $i \in \{1, \dots, h\}$ \;

    \eIf{$p \in \mathcal{P}_{\mathrm{imp}}$}{
        \For{$i \leftarrow 1$ \KwTo $h$}{
            $\mathbf{e}_{\mathrm{imp},i} \leftarrow \mathrm{Enc}(q \,\|\, \hat{T}_i)$ \;
            \tcp{per-table sub-entry}
            $\mathcal{U} \leftarrow \mathcal{U} \cup \{\mathbf{e}_{\mathrm{imp},i}\}$ \;
        }
    }{
        $\mathbf{e}_{\mathrm{exp}} \leftarrow \mathrm{Enc}(q \,\|\, \hat{T}_1 \,\|\, \cdots \,\|\, \hat{T}_h)$ \;
        \tcp{joined entry}
        $\mathcal{U} \leftarrow \mathcal{U} \cup \{\mathbf{e}_{\mathrm{exp}}\}$ \;
    }
}

\BlankLine
\Return $\mathcal{U}$

\end{algorithm}


\subsubsection{Details of Candidate Discovery}
\label{app:candidate_discovery}

As introduced in Section~\ref{para:candidate_discovery}, candidate edges $e$ are collected via three complementary signals, each capturing a distinct aspect of joinability.

\begin{itemize}
     \item \textbf{Value-Overlap ($\mathcal{P}_V$):} Using an inverted value index, we retrieve edges with Jaccard similarity over $0.10$ on sampled cell values.
    \item \textbf{Structural Pattern ($\mathcal{P}_S$):} Using an inverted index over normalized column names, we retrieve edges whose names match a FK naming convention or share an identical normalized token across different tables.
    \item \textbf{Column Similarity ($\mathcal{P}_B$):} Using the column embedding index built in Section \ref{app:index_construction}, we retrieve column edges whose similarity exceeds an adaptive threshold $\tau_{\mathrm{sim}}$, defined as the 85th percentile of corpus-wide similarities.
\end{itemize}


\subsubsection{Details of Join Path Construction}
\label{app:Path_Selection}

\paragraph{Rationale for Join Path Construction.}
As shown in Table~\ref{tab:gold-path-stats}, 2-hop and 3-hop queries account for the majority of cases across all benchmarks. Accordingly, we restrict join path generation to 2-hop and 3-hop paths. Table~\ref{tab:gold-path-stats} also highlights the diversity of path connectivity patterns. Although explicit-only paths dominate across datasets (70.2 to 88.7\%), implicit-only paths (11.3 to 21.3\%) and mixed explicit--implicit paths (up to 8.5\%) are observed. Our pipeline supports all three connectivity patterns by modeling explicit and implicit edges and combining them during join path construction.


\paragraph{Implementation Details of Path Construction.}
\algname{} generates 2-hop and 3-hop join paths from explicit and implicit edges. Candidate edges are enumerated from $\mathcal{P}_S \cup \mathcal{P}_B \cup \mathcal{P}_V$ and scored to obtain $s_{\mathrm{str}}$ and $s_{\mathrm{sem}}$ for each path $p$. Prior to Lazy Greedy selection, candidate paths are filtered using a quality threshold $\tau$ to ensure that only sufficiently reliable join edges are included in each pool.

\textbf{2-hop paths.}
Each candidate edge $e$ corresponds to a 2-hop path, and only paths satisfying $f(p) \ge \tau_{\mathrm{2h}}$ are retained: paths induced from $\mathcal{P}_S \cup \mathcal{P}_V$ form the explicit candidate pool $\mathcal{P}_{\mathrm{exp}}$, while those from $\mathcal{P}_B \cup \mathcal{P}_V$ form the implicit candidate pool $\mathcal{P}_{\mathrm{imp}}$.




\textbf{3-hop paths.}
A table sequence $(T_i, T_k, T_j)$ is formed by selecting an intermediate table $T_k$ and two of its neighbors in the join graph induced by 2-hop paths. To prevent weak edges from entering longer chains, a stricter threshold $\tau_{\mathrm{3h}} > \tau_{\mathrm{2h}}$ is applied to 3-hop paths:

\begin{equation}
    \min\bigl(f(p_{ik}),\, f(p_{kj})\bigr) \ge \tau_{\mathrm{3h}},
\end{equation}
where $p_{ik}, p_{kj} \in \mathcal{P}_{\mathrm{exp}}$ for explicit 3-hop paths and $p_{ik}, p_{kj} \in \mathcal{P}_{\mathrm{imp}}$ for implicit 3-hop paths. For each intermediate table $T_k$, we retain only the highest-scoring triple $(T_i, T_k, T_j)$.


\subsection{Details of Table Serialization}
\label{app: serialization}
Both Instructional Query Generation and Vertical Partitioning and Indexing use the same serialized representation of each table. To preserve join-relevant evidence under a limited row budget, we reorder table rows based on the value overlap identified along each join edge.

Rows whose join-column values fall within the overlap are placed first, while the remaining rows retain their original order. For a table at the center of a 3-hop path, which is incident to two join edges, we take the union of the overlap sets from both edges. Thus, a row is moved to the front if it contains a value matching either edge. We then serialize the first ten rows of the reordered table to meet the input length limits of both the query generation LLM and the encoder. By reordering the rows before applying this truncation, join-relevant rows are prioritized in the table representation.


\begin{table}[t]
\centering
\small
\begin{tabular*}{\columnwidth}{@{\extracolsep{\fill}}lrrr}
\toprule
\textbf{Metric} & \multicolumn{1}{c}{\textbf{MMQA}} & \multicolumn{1}{c}{\textbf{SPIDER}} & \multicolumn{1}{c}{\textbf{BIRD}} \\
\midrule
\multicolumn{4}{l}{\textit{\textbf{Hop Distribution}}} \\
\hspace*{1em} 1-hop           & 0.0\%  & 55.6\%  & 23.7\%  \\
\hspace*{1em} 2-hop           & 78.3\% & 38.0\%  & 61.0\%  \\
\hspace*{1em} 3-hop           & 19.7\% & 5.8\%   & 13.0\%  \\
\hspace*{1em} 4+-hop          & 2.0\%  & 0.6\%   & 2.3\%   \\
\hspace*{1em} \textbf{2-hop + 3-hop} & \textbf{98.0\%} & \textbf{43.8\%} & \textbf{74.0\%} \\
\midrule
\multicolumn{4}{l}{\textit{\textbf{Path Connectivity (Multi-hop)}}} \\
\hspace*{1em} Explicit-only   & 70.2\% & 88.7\%  & 75.8\%  \\
\hspace*{1em} Implicit-only   & 21.3\% & 11.3\%  & 20.5\%  \\
\hspace*{1em} Mixed           & 8.5\%  & 0.0\%   & 3.6\%   \\
\bottomrule
\end{tabular*}
\caption{Distribution of gold join paths across the evaluated benchmarks.}
\vspace{-0.6cm}
\label{tab:gold-path-stats}
\end{table}

\subsection{Implementation Details of Random-View Augmentation}
\label{app:random_view}

The Random-view baseline constructs an augmented corpus by replacing \algname{}'s join-aware entries with randomly projected sub-tables. For each table $T_i \in \mathcal{C}$, we randomly sample a non-empty subset of columns $c$ from its schema $S_i$:
\begin{equation}
    c \sim \mathrm{Uniform}\!\left(
    \{\, c \subseteq S_i \mid 1 \le |c| \le |S_i| \,\}
    \right).
\end{equation}
This yields a projected sub-table $\hat{T}_i = \Pi_c(T_i)$, which is encoded without any associated instructional query to produce the random-view entry:
\begin{equation}
    \mathbf{e}_{\mathrm{rand}} = \mathrm{Enc}\!\left(\hat{T}_i\right).
\end{equation}

To ensure a fair comparison, the total number of random-view entries is matched to the augmented corpus size of \algname{}:
\begin{equation}
    |\{\mathbf{e}_{\mathrm{rand}}\}| = |\{\mathbf{e}_{\mathrm{imp}}\}| + |\{\mathbf{e}_{\mathrm{exp}}\}|.
\end{equation}

\begin{table*}[t]
\vspace{-0.6cm}
\centering
\footnotesize 
\setlength{\tabcolsep}{6pt}
\resizebox{\textwidth}{!}{
\begin{tabular}{ll cc|cc|cc}
\toprule
\multirow{2}{*}{\textbf{Setting}} & \multirow{2}{*}{\textbf{Model}} 
& \multicolumn{2}{c|}{\textbf{SPIDER}} 
& \multicolumn{2}{c|}{\textbf{MMQA}} 
& \multicolumn{2}{c}{\textbf{BIRD}} \\ 
\cmidrule(lr){3-4} \cmidrule(lr){5-6} \cmidrule(lr){7-8}
& & Slot R@2 & Slot R@5 & Slot R@2 & Slot R@5 & Slot R@2 & Slot R@5 \\ 
\midrule

\multirow{9}{*}{\textbf{3-Hop}} 
& Contriever          & 56.10 & 91.10 & 38.10 & 55.50 & 45.80 & 78.70 \\
& Stella-v5            & 58.89 & 87.78 & \underline{47.30} & \underline{71.70} & \underline{52.17} & 85.33 \\ 
& JAR                  & \underline{62.80} & \underline{94.40} & 44.70 & 64.20 & \textbf{54.30} & \underline{86.20} \\
& Greedy-JAR           & 60.00 & 90.60 & 44.50 & 61.80 & \textbf{54.30} & \textbf{88.30} \\
& DTR                  & 56.67 & 89.44 & 44.39 & 66.46 & 48.33 & 80.83 \\
& MURRE                & 52.78 & 85.00 & 34.05 & 56.69 & 45.83 & 74.83 \\
& CORE-T               & 55.56 & 92.22 & 45.80 & 64.38 & 52.67 & 84.33 \\
\cmidrule(lr){2-8}
& \algname{}           & \textbf{64.44} & \textbf{100.00} 
& \textbf{49.41} & \textbf{75.09} 
& 53.83 & 85.83 \\[0.5ex]
&                      & (\lift{2.61}\%) & (\lift{5.93}\%) 
& (\lift{4.46}\%) & (\lift{4.73}\%) 
& (\drop{0.87}\%) & (\drop{2.80}\%) \\ 
\midrule

\multirow{9}{*}{\textbf{2-Hop}} 
& Contriever          & 78.20 & 97.80 & 49.10 & 63.60 & 62.00 & 84.00 \\
& Stella-v5            & 76.97 & 96.06 & 55.76 & \underline{72.63} & 64.67 & 92.29 \\ 
& JAR                  & 85.00 & \underline{98.10} & 55.90 & 65.20 & \textbf{80.30} & 92.50 \\
& Greedy-JAR           & \underline{86.50} & 95.70 & \underline{56.80} & 64.00 & \underline{79.10} & \underline{93.70} \\
& DTR                  & 80.28 & 96.18 & 52.47 & 66.37 & 62.62 & 87.38 \\
& MURRE                & 74.80 & 92.22 & 49.15 & 70.20 & 56.20 & 82.46 \\
& CORE-T               & 80.21 & 96.97 & 54.51 & 67.23 & 72.78 & 92.57 \\
\cmidrule(lr){2-8}
& \algname{}           & \textbf{86.88} & \textbf{98.73} 
& \textbf{56.87} & \textbf{72.81} 
& 68.42 & \textbf{93.90} \\[0.5ex]
&                      & (\lift{0.44}\%) & (\lift{0.64}\%) 
& (\lift{0.12}\%) & (\lift{0.25}\%) 
& (\drop{14.80}\%) & (\lift{0.21}\%) \\ 
\midrule

\multirow{9}{*}{\textbf{Unified}} 
& Contriever          & 84.80 & 97.90 & 46.30 & 61.50 & 62.80 & 84.60 \\
& Stella-v5            & 85.46 & 97.42 & 53.10 & \underline{71.72} & 66.71 & 91.36 \\ 
& JAR                  & 87.80 & \underline{98.20} & 53.10 & 64.60 & \textbf{75.10} & 90.80 \\
& Greedy-JAR           & \underline{88.20} & 97.00 & \underline{53.80} & 63.20 & \underline{74.40} & 91.70 \\
& DTR                  & 86.99 & 97.45 & 50.36 & 65.96 & 65.80 & 87.78 \\
& MURRE                & 77.27 & 92.60 & 45.62 & 66.82 & 58.84 & 82.51 \\
& CORE-T               & 87.58 & 98.22 & 52.16 & 66.13 & 74.71 & \underline{92.54} \\
\cmidrule(lr){2-8}
& \algname{}           & \textbf{88.48} & \textbf{98.75} 
& \textbf{54.26} & \textbf{72.88} 
& 69.66 & \textbf{92.73} \\[0.5ex]
&                      & (\lift{0.32}\%) & (\lift{0.54}\%) 
& (\lift{0.86}\%) & (\lift{1.62}\%) 
& (\drop{7.24}\%) & (\lift{0.21}\%) \\ 
\midrule

\multicolumn{8}{l}{\textit{Ablation of \algname{} (Unified)}} \\
\multicolumn{2}{l}{\hspace{3mm} w/o joined entry}
& 87.55 & 99.12 & 54.21 & 73.34 & 69.66 & 92.73 \\
\multicolumn{2}{l}{\hspace{7mm}{\tiny (= w/ implicit-only)}} & & & & & & \\
\multicolumn{2}{l}{\hspace{3mm} w/ explicit-only}
& 88.48 & 98.61 & 52.22 & 70.73 & 66.36 & 90.79 \\
\multicolumn{2}{l}{\hspace{3mm} Random-view}
& 84.66 & 97.31 & 52.39 & 70.80 & 66.25 & 90.78 \\
\multicolumn{2}{l}{\hspace{3mm} Random-query}
& 85.99 & 97.05 & 54.97 & 73.15 & 68.73 & 92.51 \\
\bottomrule
\end{tabular}
}
\caption{Adjusted recall (Slot R@$k$) across 3-Hop, 2-Hop, and Unified configurations. The best results are in \textbf{bold} and the second best are \underline{underlined}. $\color{red}\blacktriangle$ and $\color{blue}\blacktriangledown$ indicate improvement and degradation over the strongest baseline within each metric column.}
\vspace{-0.6cm}
\label{tab:slot}
\end{table*}


\section{Additional Evaluation}
This section provides supplementary evaluations of retrieval performance and Join Path Discovery.

\subsection{Retrieval formulation and unit-level evaluation}

\subsubsection{Adjusted recall}
\label{app: adjusted-recall}
\algname{} constructs an augmented retrieval corpus of three unit types: Original entries, Implicit sub-entries, and Explicit joined entries. Since joined entries may span multiple gold tables, they may inflate standard recall metrics. For example, given a gold set $\{A, B, C\}$, retrieving a joined unit $\{A, B, C\}$ at rank 1 results in all three tables being counted as correctly retrieved under standard R@1. While this reflects the advantage of joined entry in capturing multiple relevant tables jointly, it can also obscure the retrieval capacity of individual ranking positions.


To provide a more fine-grained evaluation, we additionally report Slot R@$k$, which re-ranks tables within the top-$k$ retrieved units based on their individual sub-table similarity. Let $\mathcal{R}_k(q) = \{u_1,\ldots, u_k\}$ be the top-$k$ retrieved corpus units sorted by similarity to query $q$, where each unit $u$ is associated with a set of tables in its join path, $\mathrm{tables}(u) = \{t_1,\ldots,t_h\}$. The candidate table set is defined as:

\begin{equation}
    \mathcal{T}_k(q) = \bigcup_{u \in \mathcal{R}_k(q)} \mathrm{tables}(u).
\end{equation}

Each table $t \in \mathcal{T}_k(q)$ is re-scored by:

\begin{equation}
    s(t, q) = \max_{\mathbf{e} \in \mathcal{P}(t)} \cos\!\left(\mathrm{Enc}(q),\, \mathbf{e}\right),
\end{equation}
where $\mathcal{P}(t)$ denotes the embeddings of all sub-table units associated with the table $t$ across join paths. 
Tables in $\mathcal{T}_k(q)$ are then re-ranked according to $s(t,q)$. Slot R@$k$ measures whether all gold tables appear within the top-$k$ re-ranked tables.


\paragraph{RQ8. Do the performance gains of \algname{} remain significant under Slot R@k evaluation?} 

As summarized in Table~\ref{tab:slot}, \algname{} consistently maintains its advantage even under more fine-grained metrics. On SPIDER and MMQA, \algname{} outperforms decomposition-based methods, achieving gains over the strongest baseline in the 3-hop configuration with relative improvements of 2.61\% and 4.46\% under Slot R@2, indicating that the benefits of \algname{} scale with increasing hop complexity, consistent with Table~\ref{tab:main_results}. On BIRD, \algname{} remains competitive with decomposition-based methods despite a modest performance drop, and we provide additional analysis in Appendix~\ref{app: BIRD}. These results suggest that the gains primarily arise from noise reduction and inter-table relationship modeling enabled by \algname{}'s join-aware augmentation, rather than inflated coverage from joined entries.


\subsubsection{Design rationale for max-pooling}
\algname{} does not enforce that retrieved tables originate from a single coherent join path. This is a deliberate design choice, as the objective of multi-hop table retrieval is to recover the complete set of gold tables within the top-k results rather than a single reasoning chain. Under this objective, max-pooling assigns each table the score of its best matching indexed view. Since a table may participate in multiple join paths, aggregating scores across all indexed views would dilute relevant evidence with unrelated contexts.

For the same reason, enforcing path coherence does not necessarily improve robustness to false joins. If an entire path were retrieved as a single unit, a single incorrect edge could compromise the whole chain. In contrast, \algname{} evaluates augmented units independently and retains only the highest-scoring unit for each table, limiting the influence of spurious paths.


\subsubsection{Design rationale for asymmetric encoding}
\label{app: asymmetric}

The asymmetric encoding reflects the distinct retrieval signals of explicit and implicit paths. For explicit paths, the structural join itself is the primary information, making a joined representation effective (equation 7 in Section ~\ref{subsec: Vertical Partitoning}). In contrast, implicit paths lack deterministic relational joins, so forcing a unified representation would introduce arbitrary structure. \algname{} therefore encodes each sub-table independently (equation 6 in Section ~\ref{subsec: Vertical Partitoning}), while the shared generated query provides the semantic context linking them.

To validate this design choice, we evaluate an alternative that encodes implicit paths as joined entries (Table~\ref{tab:joined_implicit}). Although this variant improves standard R@2, it consistently underperforms under Slot R@2. This indicates that the apparent improvement is largely driven by a single joined entry covering multiple gold tables simultaneously, as discussed in Appendix~\ref{app: adjusted-recall}. Thus, the separate encoding strategy better aligns with the retrieval objective, while joined representations may still be preferable for applications that prioritize relational coherence.


\begin{table}[t]
\vspace{-0.6cm}
\centering
\small
\setlength{\tabcolsep}{5.5pt}
\begin{tabular}{@{}p{1.9cm}ccc@{}}
\toprule
 & SPIDER & MMQA & BIRD \\
 & R@2/Slot & R@2/Slot & R@2/Slot \\
\midrule
\algname{} & 91.25/\textbf{88.48} & 58.84/\textbf{54.26} & 73.64/\textbf{69.66} \\
Implicit joined & \textbf{92.29}/87.61 & \textbf{61.57}/53.68 & \textbf{74.53}/67.25 \\
\bottomrule
\end{tabular}
\vspace{-0.2cm}
\caption{Evaluation of the asymmetric design under the unified setting, where implicit paths are encoded as joined entries.}
\vspace{-0.6cm}
\label{tab:joined_implicit}
\end{table}
\begin{table*}[t]
\vspace{-0.5cm}
\centering
\small
\renewcommand{\arraystretch}{1.05}
\begin{tabular*}{\textwidth}{@{\extracolsep{\fill}}lccccccc@{}}
\toprule
\multicolumn{8}{@{}l}{\textbf{(a) Gold-path agreement}} \\
\midrule
Dataset & Precision & Recall & F1
& \multicolumn{2}{c}{PEARL} & \multicolumn{2}{c}{Oracle Join} \\
\cmidrule(lr){5-6} \cmidrule(lr){7-8}
& & & & R@2 & R@5 & R@2 & R@5 \\
\midrule
SPIDER & 0.58 & 0.65 & 0.61 & \textbf{91.25} & \textbf{99.33} & 88.09 & 97.24 \\
MMQA   & 0.40 & 0.36 & 0.38 & 58.84 & 76.21 & \textbf{61.68} & \textbf{85.51} \\
BIRD   & 0.58 & 0.30 & 0.39 & 73.64 & 93.36 & \textbf{84.44} & \textbf{96.27} \\
\bottomrule
\end{tabular*}

\vspace{0.8em}

\begin{tabular*}{\textwidth}{@{\extracolsep{\fill}}lcccccc@{}}
\toprule
\multicolumn{7}{@{}l}{\textbf{(b) Sensitivity to join path quality}} \\
\midrule
& \multicolumn{2}{c}{SPIDER} & \multicolumn{2}{c}{MMQA} & \multicolumn{2}{c}{BIRD} \\
\cmidrule(lr){2-3} \cmidrule(lr){4-5} \cmidrule(lr){6-7}
Method & R@2 & R@5 & R@2 & R@5 & R@2 & R@5 \\
\midrule
PEARL & \textbf{91.25} & \textbf{99.33} & \textbf{58.84} & \textbf{76.21} & \textbf{73.64} & \textbf{93.36} \\
\quad +50\% noise  & 89.28 & 97.53 & 55.03 & 72.98 & 69.03 & 89.44 \\
\quad +100\% noise & 83.56 & 96.38 & 51.72 & 70.02 & 64.41 & 87.92 \\
Stella-v5          & 85.46 & 97.42 & 53.10 & 71.72 & 66.71 & 91.36 \\
\midrule
Implicit Match (\%) & \multicolumn{2}{c}{62} & \multicolumn{2}{c}{78} & \multicolumn{2}{c}{69} \\
\bottomrule
\end{tabular*}
\caption{Evaluation of join path discovery.
(a) Gold-path agreement and retrieval performance using Join Path Discovery (PEARL) versus Oracle Join.
(b) Sensitivity of retrieval performance to join path corruption and the contribution of implicit join paths.}
\label{tab:join-path-analysis}
\vspace{-0.3cm}
\end{table*}
\begin{table}[t]
\centering
\footnotesize
\begin{tabular*}{\columnwidth}{@{\extracolsep{\fill}}lccc@{}}
\toprule
Dataset & Median Lift & NN sim. (2h / 3h) & Dup. (\%) \\
\midrule
SPIDER & +0.200 & 0.713 / 0.733 & 0.4 \\
MMQA   & +0.216 & 0.704 / 0.718 & 0.3 \\
BIRD   & +0.173 & 0.682 / 0.688 & 0.1 \\
\bottomrule
\end{tabular*}
\caption{Coverage statistics of generated queries against the random-match. NN sim.\ denotes mean nearest similarity for 2-hop / 3-hop queries.}
\label{tab:coverage}
\vspace{-0.5cm}
\end{table}


\subsection{Evaluation on Join Path Discovery and Query Coverage}
Section~\ref{sec: Ablation study} examined the evaluation of joined entries. Building on this analysis, we now shift our focus to the join paths themselves and investigate three remaining research questions.


\paragraph{RQ9. Does agreement with gold paths accurately reflect join path quality?}

Against the gold paths, Join Path Discovery achieves F1 scores of 0.61 on SPIDER, 0.38 on MMQA, and 0.39 on BIRD, with the corresponding precision and recall reported in Table~\ref{tab:join-path-analysis}. This modest agreement reflects the different objectives of gold paths and Join Path Discovery. A gold path captures the path required for an observed query, whereas Join Path Discovery identifies a broader set of paths without access to future queries. Under this objective, paths outside the gold set are not necessarily errors and may cover query intents absent from the benchmark.

To examine whether these additional paths provide retrieval value, we compare \algname{} with an oracle that indexes only gold paths. On SPIDER, \algname{} outperforms the oracle, showing that gold paths are not necessarily retrieval optimal, and that low gold agreement does not imply low retrieval value. On MMQA and BIRD, the oracle performs better, indicating room for further gains rather than a limitation of the retrieval formulation itself (consistent with the analysis in Appendix~\ref{app: BIRD}).


\paragraph{RQ10. How sensitive is retrieval performance to variations in join path quality?}
Since RQ4 shows that corpus expansion alone does not explain \algname{}’s gains, we test its dependence on join quality by replacing discovered paths with randomly constructed non-joinable paths while holding corpus size fixed. Performance decreases monotonically with the corruption rate and falls below the base encoder at 100\% settings. At 50\% corruption, \algname{} still outperforms the base encoder on all three benchmarks, as the max-pooling strategy rarely selects low-similarity spurious units.

Table~\ref{tab:join-path-analysis} further shows that this dependence on meaningful joins is also reflected in implicit paths. Their match share, the fraction of test queries whose nearest generated query originates from an implicit path, exceeds their prevalence in the generated query pool across benchmarks, particularly on BIRD. This suggests that the discovered joins align with real query intents rather than benefiting from volume. Moreover, a single Join Path Discovery pipeline captures transferable join structure across benchmarks without per-dataset tuning.

\paragraph{RQ11. Do generated queries capture real query intents without reproducing the original queries?}

RQ9 shows that low gold-path agreement does not imply low retrieval value, but it does not address whether the generated queries align with real information needs. We therefore conduct a query-level coverage analysis by measuring the cosine similarity between each test query and its nearest generated query. To provide a reference, we compare these similarities against a random-match, where each test query is paired with a randomly selected generated query.

As summarized in Table~\ref{tab:coverage}, coverage lies well above random, with the median exceeding the random-match by 0.173–0.216 across all benchmarks. The nearest-neighbor similarity is also higher for 3-hop than for 2-hop queries, consistent with the hop-wise retrieval gains observed in RQ2. Fewer than 0.4\% of the nearest generated queries have a cosine similarity of at least 0.9. Since query generation relies only on table structure and sampled rows, this low duplication suggests that the retrieval gains arise from the augmented indexing units rather than from reproducing benchmark queries.


\begin{table*}[t]
\centering
\footnotesize
\setlength{\tabcolsep}{6pt} 
\resizebox{\textwidth}{!}{
\begin{tabular}{ll cc|cc|cc}
\toprule
\multirow{2}{*}{\textbf{Setting}} & \multirow{2}{*}{\textbf{Model}}
  & \multicolumn{2}{c|}{\textbf{SPIDER}}
  & \multicolumn{2}{c|}{\textbf{MMQA}}
  & \multicolumn{2}{c}{\textbf{BIRD}} \\
\cmidrule(lr){3-4}\cmidrule(lr){5-6}\cmidrule(lr){7-8}
 & & R@2 & R@5 & R@2 & R@5 & R@2 & R@5 \\ \midrule

\multirow{5}{*}{\textbf{3-Hop}}
 & Stella-v5  & 58.89 & 87.78 & \underline{47.30} & \underline{71.70} & 52.17 & 85.33 \\
 & GJAR       & 60.60 & 91.10 & 44.50 & 61.30 & 55.70 & \textbf{88.50} \\
 & JAR        & \underline{63.30} & \underline{95.60} & 44.40 & 63.50 & \underline{56.20} & 84.80 \\
\cmidrule(lr){2-8}
 & PEARL      & \textbf{74.44} & \textbf{98.89} & \textbf{57.06} & \textbf{79.46} & \textbf{56.50} & \underline{87.67} \\[0.5ex]
 &            & (\lift{17.60}\%) & (\lift{3.44}\%) & (\lift{20.63}\%) & (\lift{10.82}\%) & (\lift{0.53}\%) & (\drop{0.94}\%) \\
\midrule

\multirow{5}{*}{\textbf{2-Hop}}
 & Stella-v5  & 76.97 & 96.06 & 55.76 & \underline{72.63} & 64.67 & 92.29 \\
 & GJAR       & \underline{86.30} & 95.80 & \underline{56.40} & 63.90 & \textbf{79.40} & \underline{93.50} \\
 & JAR        & 84.60 & \underline{97.80} & 55.50 & 64.90 & \underline{79.30} & 92.10 \\
\cmidrule(lr){2-8}
 & PEARL      & \textbf{87.02} & \textbf{99.49} & \textbf{58.92} & \textbf{75.66} & 72.16 & \textbf{94.11} \\[0.5ex]
 &            & (\lift{0.83}\%) & (\lift{1.73}\%) & (\lift{4.47}\%) & (\lift{4.17}\%) & (\drop{9.12}\%) & (\lift{0.65}\%) \\
\midrule

\multirow{5}{*}{\textbf{Unified}}
 & Stella-v5  & 85.46 & 97.42 & 53.10 & \underline{71.72} & 66.71 & 91.36 \\
 & GJAR       & \underline{88.10} & 97.10 & \underline{53.30} & 62.80 & \textbf{74.70} & \underline{91.50} \\
 & JAR        & 87.70 & \underline{98.10} & 52.60 & 64.10 & \textbf{74.70} & 90.30 \\
\cmidrule(lr){2-8}
 & PEARL      & \textbf{90.50} & \textbf{99.26} & \textbf{57.58} & \textbf{75.62} & \underline{72.55} & \textbf{93.24} \\[0.5ex]
 &            & (\lift{2.72}\%) & (\lift{1.18}\%) & (\lift{8.03}\%) & (\lift{5.44}\%) & (\drop{2.88}\%) & (\lift{1.90}\%) \\
\bottomrule
\end{tabular}
}
\caption{Retrieval performance comparison with the same LLM backbone (GPT-4o-mini). The best results are in \textbf{bold} and the second best are \underline{underlined}. The symbols $\color{red} \blacktriangle$ and $\color{blue} \blacktriangledown$ indicate the relative improvement and degradation over the strongest baseline within each metric column, respectively.}
\label{tab:ablation_4omini}
\end{table*}


\section{Further Analysis}
\label{app: Further}
In this section, we present additional experimental results to provide deeper insights into \algname{}.


\begin{table}[t]
\vspace{-0.1cm}
\centering
\small
\renewcommand{\arraystretch}{1.05}
\setlength{\tabcolsep}{3.5pt}

\begin{tabular}{lcccccc}
\toprule
& \multicolumn{2}{c}{SPIDER}
& \multicolumn{2}{c}{MMQA}
& \multicolumn{2}{c}{BIRD} \\
\cmidrule(lr){2-3}
\cmidrule(lr){4-5}
\cmidrule(lr){6-7}

LLM Backbone
& R@2 & R@5
& R@2 & R@5
& R@2 & R@5 \\
\midrule

Llama 3.3 70B
& \textbf{91.25} & \textbf{99.33}
& \textbf{58.84} & \textbf{76.21}
& \textbf{73.64} & \textbf{93.36} \\

GPT-4o mini
& 90.50 & 99.26
& 57.58 & 75.62
& 72.55 & 93.24 \\

Qwen 2.5 7B
& 88.88 & 98.97
& 56.90 & 74.81
& 69.67 & 92.39 \\

\bottomrule
\end{tabular}
\caption{Retrieval performance of PEARL with different LLM backbones.}
\label{tab:llm_backbone}
\vspace{-0.3cm}
\end{table}

\subsection{Effect of LLM Backbone}
\label{app:unified_llm}
\paragraph{Comparison under an identical backbone}
Table~\ref{tab:ablation_4omini} compares multi-hop table retrieval methods using the same LLM backbone (GPT-4o mini), isolating the effect of retrieval architecture from differences in model capability. Across most settings, \algname{} consistently outperforms competing baselines. In particular, on SPIDER and MMQA, \algname{} achieves the highest R@2 in the majority of evaluations, with gains up to +17.6\% and +20.63\% on 3-hop SPIDER and MMQA, respectively. On BIRD, \algname{} attains the best performance in R@5. These results indicate that the performance gain of \algname{} primarily arises from its retrieval design rather than the strength of the underlying backbone model.

\paragraph{Robustness across LLM Backbones}

Although \algname{} uses a 70B backbone in the main experiments, we additionally evaluate Qwen 2.5 7B Instruct and GPT-4o mini in a zero-shot setting to assess whether the method remains effective with lightweight LLMs for offline indexing.

As shown in Table~\ref{tab:llm_backbone}, \algname{} is largely insensitive to backbone capacity. Under the unified setting, the smaller models (GPT-4o mini and Qwen 2.5 7B) achieve retrieval performance comparable to the larger Llama 3.3 70B model. This robustness makes \algname{} practical when computational resources are limited, since the LLM is invoked only once during offline indexing, allowing smaller backbones to reduce the indexing cost without sacrificing retrieval effectiveness. Future improvements in lightweight LLMs can therefore be adopted directly to further reduce indexing cost without modifying the method. Together with the controlled comparison above, these results suggest that \algname{}’s gains stem from its join-aware indexing strategy rather than backbone capacity.

\subsection{Analysis of BIRD dataset}
\label{app: BIRD}

\algname{} trails GJAR and JAR on BIRD 2-hop R@2, while outperforming both on SPIDER and MMQA. To better understand this gap, we analyze the characteristics of the BIRD dataset. 


\begin{figure}[t]
  \centering
  \includegraphics[width=\columnwidth]{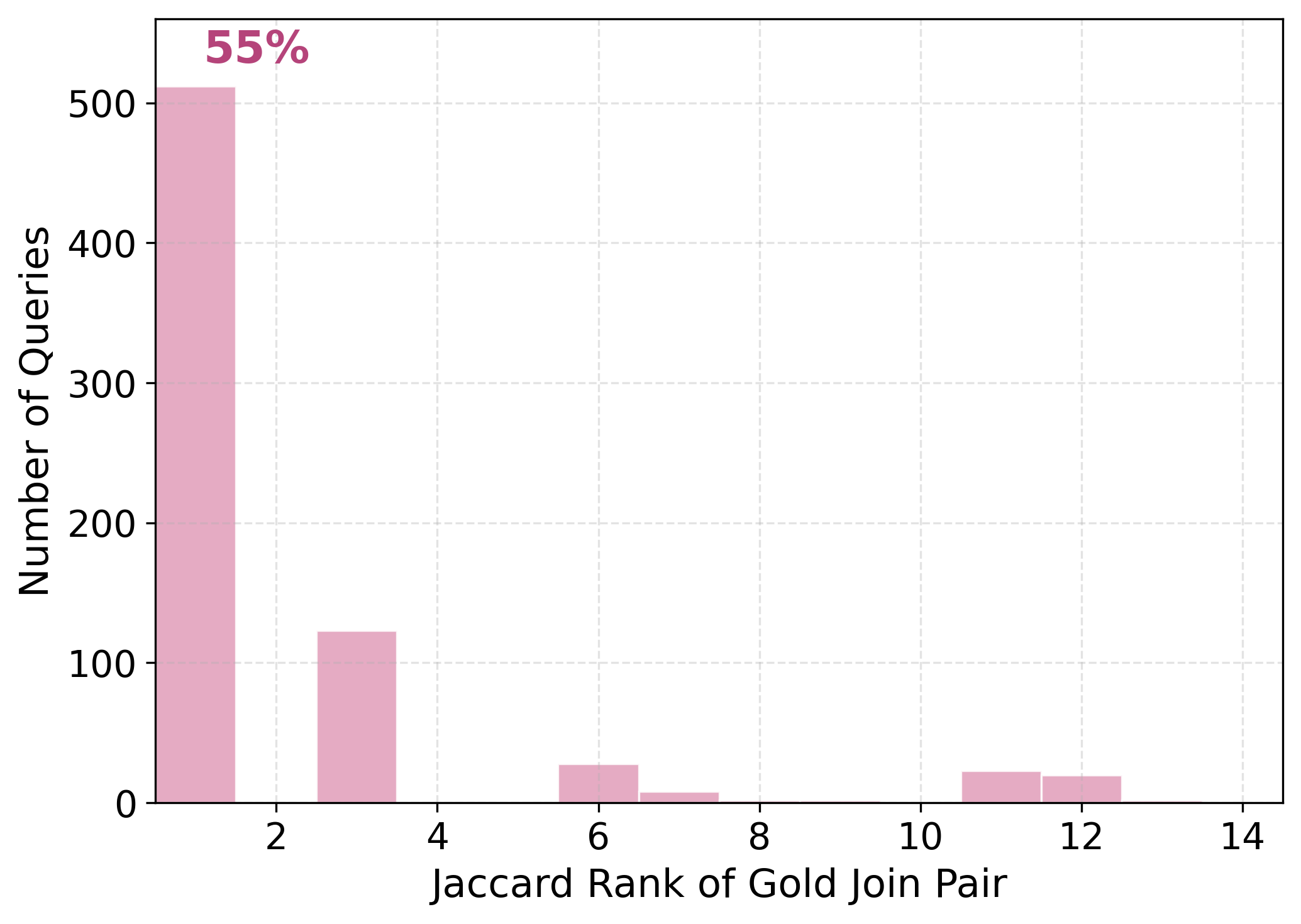}
  \vspace{-0.4cm}
  \caption{Jaccard rank distribution of gold join pairs among intra-DB table pairs for BIRD 2-hop queries.}
  \vspace{-0.2cm}
  \label{fig:bird_rank}
\end{figure}
\begin{table}[t]
\centering
\small
\setlength{\tabcolsep}{3.5pt}
\begin{tabular}{@{}lrrrrr@{}}
\toprule
& \textbf{Tables} & \textbf{Impl.} & \textbf{Expl.} & \textbf{Expand} & \textbf{Index (MB)} \\
\midrule
SPIDER & 81  & 73  & 40  & 2.40$\times$ & 0.79 \\
MMQA   & 555 & 496 & 277 & 2.39$\times$ & 5.44 \\
BIRD   & 75  & 68  & 37  & 2.40$\times$ & 0.74 \\
\bottomrule
\end{tabular}
\caption{Storage overhead introduced by \algname{}. ``Impl.'' and ``Expl.'' denote the numbers of additional implicit and explicit indexing units, respectively.}
\vspace{-0.5cm}
\label{tab:overhead}
\end{table}

\begin{table}[t]
\centering
\small
\begin{tabular}{lcccc}
\toprule
\textbf{MMQA} & \textbf{Paths} & \textbf{Emb.} & \textbf{R@2} & \textbf{R@5} \\
\midrule
Full re-indexing & 554 & 13{,}478 & 58.84 & 76.21 \\
Incremental & 244$^{\dagger}$ & 2{,}730 & 58.54 & 76.15 \\
\bottomrule
\end{tabular}
\caption{Indexing workload and retrieval performance under a
20\% corpus expansion on MMQA. $^{\dagger}$310 reused paths yield 554
indexed paths in total.}
\label{tab:incremental}
\vspace{-0.6cm}
\end{table}

As shown in Figure~\ref{fig:bird_rank}, 55\% of gold join pairs already rank first based on Jaccard similarity among all intra-database table pairs. This is partly due to the small scale of the BIRD development set, which contains only 75 tables, making offline Jaccard pre-computation over all table pairs trivial and yielding highly reliable statistical similarity for identifying gold join pairs, a setting where JAR and Greedy-JAR naturally excel. In contrast, such strong statistical priors are less likely to hold in large-scale settings, where value distributions are more diverse and similarity-based signals alone are insufficient to reliably identify correct joins.

Given the gap between real-world data and the BIRD dataset, these results can be interpreted from a different perspective. \algname{} still improves over the Stella-v5 baseline in slot-level recall on BIRD, indicating that join-aware offline indexing remains effective even under strong statistical priors.

These observations suggest that BIRD may underestimate the benefit of our approach compared to more realistic large-scale settings. This is further supported by stronger performance on MMQA, which better reflects complex multi-hop retrieval scenarios than BIRD and SPIDER~\cite{mmqa}, including consistent gains under adjusted recall and over decomposition-based baselines. We leave a more detailed investigation of large-scale database settings to future work.


\begin{figure*}[hbt!]
  \includegraphics[width=1\linewidth]{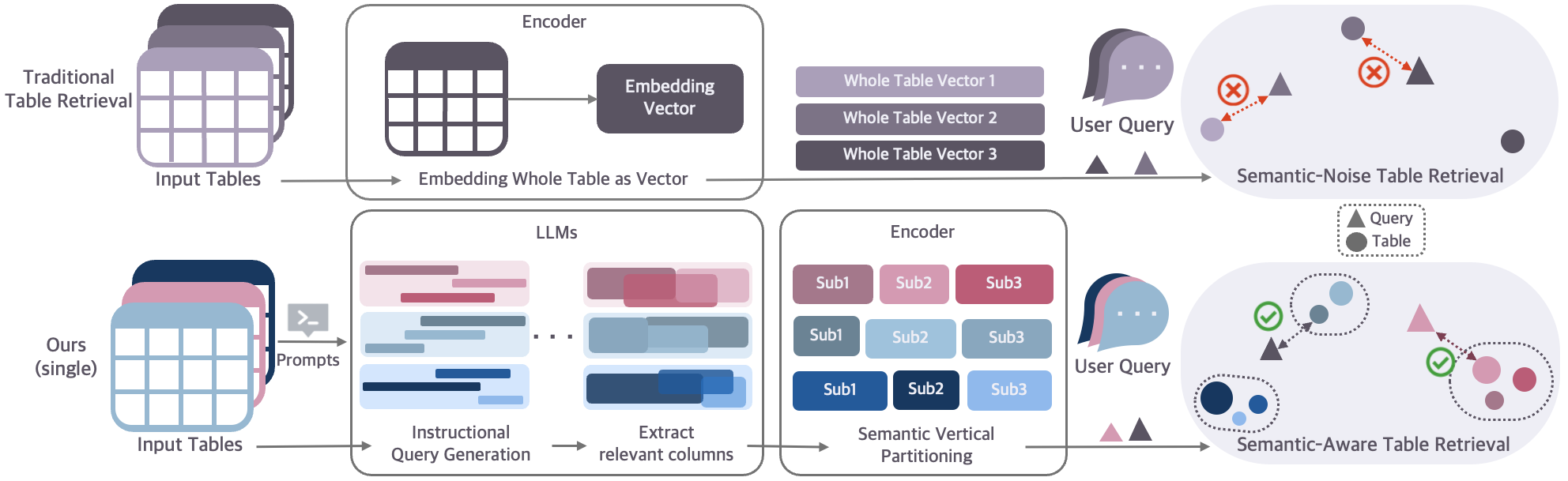} \hfill
  \vspace{-0.4cm}
  \caption {Overall procedure of \algname{} for single-hop setting.}
  \vspace{-0.3cm}
  \label{fig:singlehop_framework}
\end{figure*}


\subsection{Storage and Index Overhead}
\label{app: overhead}

While Table ~\ref{tab:efficiency_combined} reports the overall indexing efficiency, this section provides a more detailed analysis of the resulting storage and indexing overhead.

Table ~\ref{tab:overhead} additionally reports the storage and unit-count breakdown at $K=\lfloor|\mathcal{C}|/2\rfloor$. Because $K$ scales with $|\mathcal{C}|$, the corpus expansion factor remains nearly constant at approximately 2.4 times across all three benchmarks. Even with this expansion, the largest benchmark, MMQA, requires only a 5.44 MB index, roughly 3 MB larger than whole-table indexing. Moreover, the additional indexing units have little impact on online retrieval efficiency, with ANN search remaining below 32 ms even on MMQA.

This sub-linear scaling persists as the corpus budget increases.
Raising $K$ from $0.25|\mathcal{C}|$ to $1.00|\mathcal{C}|$ increases
online latency by less than a factor of two across all three
benchmarks (Table~\ref{tab:efficiency_combined}). Thus, the overhead of
corpus augmentation is incurred primarily in storage, which scales
predictably with $|\mathcal{C}|$, rather than in online latency,
consistent with \algname{}'s front-loading design. When storage is constrained, this overhead can be further reduced by lowering the corpus budget $K$, as shown in RQ6 (Section~\ref{sec:  Sensitivity Analysis}).

\subsection{Incremental Indexing under Corpus Expansion}
\label{app: dynamic}

\algname{} front-loads LLM reasoning into an offline index, raising the question of how the index can be maintained as the corpus evolves. Since \algname{} constructs join paths from pairwise join edges, updates can be localized at the edge level. Adding a table requires computing candidate join edges only for table pairs involving the new table, while existing edges remain unchanged.

Corpus units are then generated only for paths containing a new edge, with previously generated queries and embeddings reused. Similarly, removing a table only invalidates its incident edges and the corpus units associated with paths containing those edges. Thus, the update cost depends primarily on the affected edges and paths rather than requiring the entire corpus to be re-indexed.

We empirically evaluate the addition case on MMQA by splitting its 555 tables into an initial set of 444 tables (80\%) and an update set of 111 tables (20\%). Full re-indexing rebuilds the index from scratch over all 555 tables, rediscovering all join edges and regenerating all queries. In contrast, incremental indexing starts from the index built on the initial 444 tables and performs Join Path Discovery only for table pairs involving at least one newly added table. All previously generated queries and embeddings are reused unchanged. Both settings are evaluated on the same test set. As shown in Table~\ref{tab:incremental}, incremental indexing generates queries for 244 paths while reusing 310 paths from the existing index. This requires only 44\% of the LLM query generation cost of full re-indexing while achieving nearly the same retrieval performance, indicating that the statistics inherited from the existing index remain reliable under a 20\% corpus expansion. \algname{} can therefore accommodate periodic corpus expansion without rebuilding the entire index.


\subsection{Toward Adaptive Offline-Online Indexing}
\label{app: adaptive}

\algname{} generates all reasoning units offline before deployment, prioritizing low online latency and eliminating LLM inference from the retrieval pipeline. However, our results suggest that fully offline indexing may not always be necessary. RQ6 (Section~\ref{sec: Ablation study}) shows that retrieval gains are concentrated in a small subset of high-scoring join paths. Figure~\ref{fig:efficiency} further shows that offline indexing becomes advantageous only after its preprocessing cost of offline indexing is amortized, reaching the break-even point after 22–35\% of the query workload. Together, these observations suggest an adaptive indexing strategy that precomputes only the join paths whose expected retrieval benefit outweighs their indexing cost while resolving the remaining paths online when needed.

\begin{table*}[t]
\vspace{-0.5cm}
\centering
\footnotesize 
\setlength{\tabcolsep}{0pt} 
\begin{tabular*}{\textwidth}{@{\extracolsep{\fill}} l cccccc}
\toprule
\multicolumn{1}{c}{\multirow{2}{*}{\textbf{Model}}} & \multicolumn{3}{c}{\textbf{FeTaQA}} & \multicolumn{3}{c}{\textbf{FeTaQA ($\ge$ 11 Cols)}} \\ \cmidrule(lr){2-4} \cmidrule(lr){5-7} 
 & R@1 & R@5 & R@10 & R@1 & R@5 & R@10 \\ \midrule

BM25                 & 20.33 & 31.82 & 36.43 & 8.77  & 15.80 & 19.88 \\
TF-IDF               & 8.32  & 17.32 & 22.07 & 10.30 & 18.86 & 23.24 \\
SPLADE               & 29.21 & 46.59 & 53.75 & 14.78 & 33.03 & 41.69 \\ 
DTR                  & 18.03 & 35.56 & 44.01 & 13.25 & 26.81 & 36.60 \\
TAPAS                & 9.07  & 19.00 & 25.14 & 6.63  & 17.33 & 23.34 \\ \midrule

Stella-v5 (baseline) & \underline{35.15} & \underline{53.85} & \underline{61.75} & \underline{20.18} & \underline{38.63} & \underline{48.32} \\ 
\algname{}           & \textbf{40.69} & \textbf{59.89} & \textbf{67.03} & \textbf{27.83} & \textbf{46.69} & \textbf{55.35} \\[0.5ex]
                     & (\lift{15.8}\%) & (\lift{11.2}\%) & (\lift{8.6}\%) & (\lift{37.9}\%) & (\lift{20.9}\%) & (\lift{14.5}\%) \\ \midrule

\multicolumn{7}{l}{\textit{Ablation of \algname{}}} \\
\hspace{3mm} w/ subtable only    & 33.37 & 51.67 & 58.44 & 18.96 & 36.90 & 46.08 \\
\hspace{3mm} w/ query only       & 34.38 & 52.73 & 60.04 & 22.94 & 41.18 & 48.52 \\
\hspace{3mm} w/ query \& header  & 35.01 & 53.46 & 60.61 & 23.24 & 41.90 & 49.95 \\
\bottomrule
\end{tabular*}
\vspace{1mm}
\vspace{-0.1cm}
\caption{Overall performance results on single-hop setting. The best results are in \textbf{bold} and the second best are \underline{underlined}. The symbol $\color{red} \blacktriangle$ indicates the relative improvement over the Stella-v5 baseline.}
\vspace{-0.3cm}
\label{tab:single_performance}
\end{table*}


\section{Single-Hop Retrieval Setting}
\label{app: quocca}

\algname{} also generalizes to single-hop retrieval scenarios. The overall pipeline remains identical to the main framework(Figure~\ref{fig:framework}), except that it omits multi-table Join Path Discovery. Instead, \algname{} is inspired by Random-view variants, and augments individual tables by generating instructional queries that capture potential user intents, incorporating them into semantically enriched table representations. 

Figure~\ref{fig:singlehop_framework} illustrates the overview of \algname{} in single-hop setting. In the offline indexing stage, \algname{} first decomposes
input tables into sub-tables by generating instructional queries that capture potential user intents. These queries are then incorporated into the sub-tables to construct a semantically augmented embedding corpus, offering diverse views of the original tables. In the online retrieval stage, \algname{} calculates the relevance score between a given query and a table by pooling semantically aligned
views of the table.


\subsection{\textbf{Datasets and Evaluation Metrics}}
We evaluate \algname{} on the single-hop table benchmark FeTaQA~\cite{fetaqa}, which contains complex schemas. Since our method does not require training, we treat the entire dataset as a unified retrieval pool without introducing any supervision leakage. This setup yields a more practical search space that better reflects real-world retrieval scenarios. We report standard Recall@$k$.


\subsection{\textbf{Baselines and Implementation Details}}
We compare \algname{} with diverse baselines, categorized into three paradigms:
(1) \emph{Sparse retrievers}  BM25, TF-IDF, and SPLADE~\cite{SPLADE};
(2) \emph{Structure-aware dense retrievers} TAPAS~\cite{TAPAS} and DTR~\cite{DTR}, which explicitly model tabular structures and their underlying relational logic; and
(3) \emph{Text-based dense retrievers} Stella-v5~\cite{stella}, a high-performance pretrained encoder that has demonstrated superior table-to-text alignment on TARGET benchmark~\cite{target}.
For architectural consistency, \algname{} adopts Stella-v5 as the backbone encoder and Llama 3.3 70B for query generation. All baselines and \algname{} are evaluated under a zero-shot setting with 512 tokens.

\subsection{Experimental Results}
With Table \ref{tab:single_performance}, we answer the following research questions to analyze the effectiveness of \algname{} on single-hop retrieval quality.

\textbf{RQ1. Does vertical partitioning improve single-hop retrieval accuracy?} 

\algname{} consistently outperforms all baselines in R@1 and R@5 on FeTaQA benchmark. To further analyze these gains, we evaluate the FeTaQA($\ge 11$ Cols) subsets, where \algname{} achieves a 20.9\% (8.1\%p) gain in R@5 over the strongest baseline. The larger improvement on wide tables suggests that vertical partitioning effectively mitigates information dilution by pruning irrelevant columns, particularly in tables with many attributes that introduce substantial retrieval noise. By identifying query-relevant contexts in advance, \algname{} constructs semantically salient sub-table fragments that better align with downstream queries.


\textbf{RQ2. Are both sub-tables and augmented queries effective for retrieval?}
We conduct an ablation study of \algname{} to evaluate the contribution of each component in representing table views. \algname{} consistently outperforms variants utilizing only sub-tables, augmented queries, or query-header combinations. 

Notably, the performance gain by utilizing both query and sub-tables persists across wide-table and multi-hop scenarios. These results demonstrate the synergy between vertically partitioned tables and query-conditioned encoding in effectively filtering irrelevant semantic noise.

\begin{figure}[t]
  \centering
  \includegraphics[width=\columnwidth]{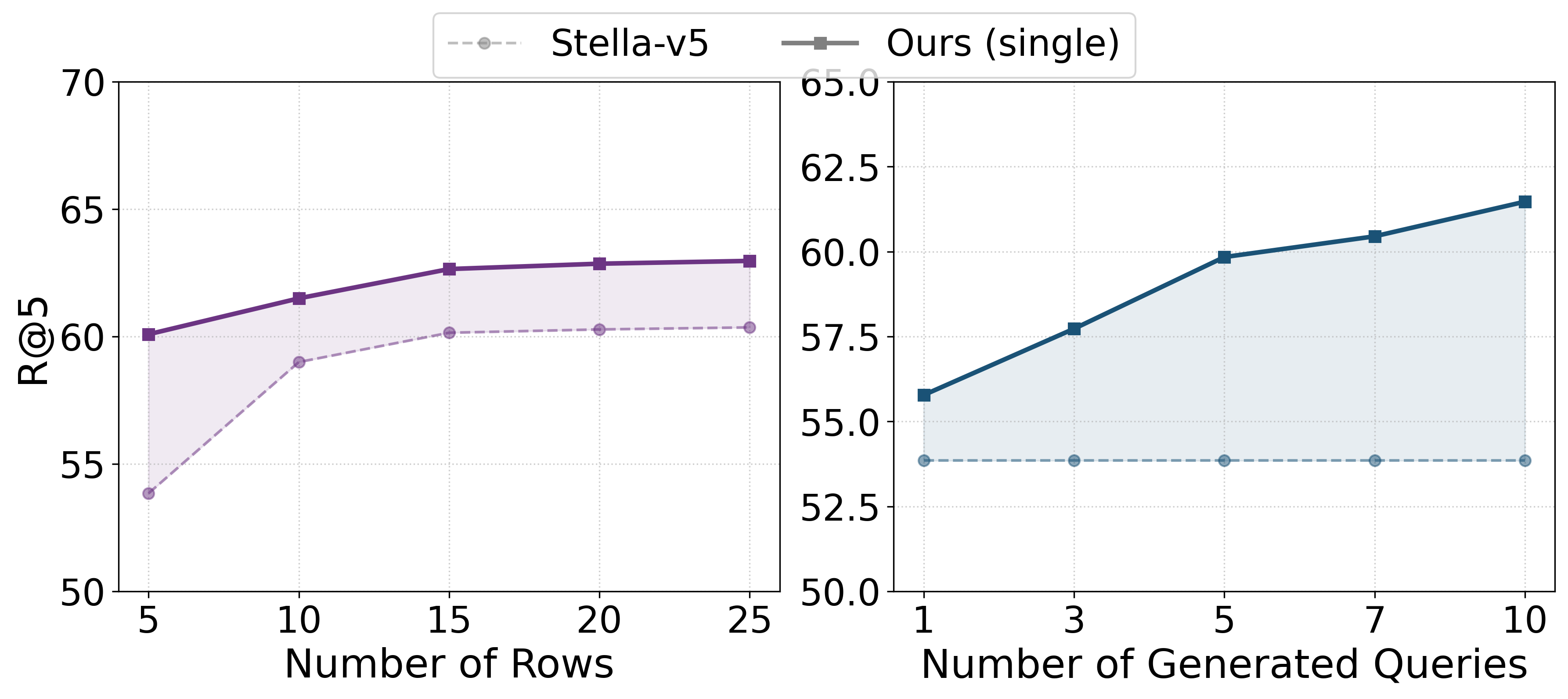}
  \caption{ Retrieval performance (R@5) over different hyperparameters of \algname{} on the FeTaQA dataset.}
  \label{fig:single_sensitivity}
\end{figure}

\textbf{RQ3. How do the number of queries $N$ affect the effectiveness of \algname{}?}
To investigate the impact of key hyperparameters in \algname{} in single-hop setting, we perform a sensitivity analysis on the number of generated queries. As shown in Figure \ref{fig:single_sensitivity}, R@5 with $N=1$ already outperforms the baseline, with performance consistently improving as $N$ increases. This suggests that the diversity of instructional queries is positively correlated with the coverage of potential user intents, highlighting an opportunity for further scaling the benefit of \algname{}. Regarding $k_\text{avg}$, retrieval performance saturates between 3 and 5, validating our guidance for setting the column subset size.


\section{Prompt Templates}
\label{appendix:prompt}
This section presents prompt templates for instructional query generation in \algname{}. All prompts use the LLM configuration described in Section~\ref{subsec: Query Generation}.
\begin{figure*}[t]
\small
\centering
\begin{promptbox}{Explicit Join Prompt}
\label{fig:explicit_prompt}
\ttfamily\setlength{\parindent}{0pt}

\textbf{[SYSTEM PROMPT]} \\
You are an expert in "Schema-Aware Multi-hop Database Question Generation".\\
Your goal is to generate ONE natural but complex question that acts as a unique signature for the explicit structural chain connecting these \{n\_tables\} tables.

\vspace{0.5em}
\#\#\# CONTEXT: EXPLICIT JOIN SEMANTICS \\
These \{n\_tables\} tables are strictly connected by explicit foreign key (FK) / primary key (PK) relationships.
\begin{itemize}[leftmargin=1.5em, noitemsep, topsep=2pt]
    \item \textbf{JOIN concept bridge:} The INPUT block spells out this structural chain; your question MUST reflect that same bridge in natural language (each hop is a concrete semantic link, not vague "relatedness").
    \item \textbf{Stepwise necessity:} The answer MUST require following EVERY join link in order—one hop at a time. Skipping any table breaks the query's meaning or makes it unanswerable.
    \item \textbf{Traversal direction:} You MAY anchor at either end (Table 1 or Table \{n\_tables\}). The solver may reason forward along the chain or reverse along the same links (as in the INPUT); do not invent shortcuts or alternate join paths.
\end{itemize}

\vspace{0.5em}
\#\#\# STRATEGY: JOIN-AWARE ANCHORING \& TRAVERSAL
\begin{enumerate}[leftmargin=1.5em, noitemsep, topsep=2pt]
    \item \textbf{Semantic Translation:} Understand \textit{why} two tables are joined based on their column names (e.g., \texttt{artist.id = album.artist\_id} $\rightarrow$ "Artist released Album"). Translate these structural links into natural language relations (verbs, ownership).
    \item \textbf{Anchor:} Start at one end of the chain (Table 1 or Table \{n\_tables\}) using specific proper noun(s) from the provided data rows.
    \item \textbf{Traverse via Relations:} Force the solver to follow the semantic path (e.g., "Find the manager (Table 3) of the department (Table 2) where employee X (Table 1) works").
    \item \textbf{Target:} Retrieve the specific attribute at the far end of the chain.
    \item \textbf{Discriminative wording:} Do NOT use pronouns (he, she, it, they); use explicit proper nouns from the rows.
\end{enumerate}

\vspace{0.5em}
\#\#\# RULES
\begin{enumerate}[leftmargin=1.5em, noitemsep, topsep=2pt]
    \item Output ONLY ONE valid JSON object inside the schema defined below. Output ONLY the JSON object. No markdown code fences. No extra text.
    \item The question MUST:
    \begin{itemize}[leftmargin=1.em, noitemsep, topsep=1pt]
        \item Require ALL \{n\_tables\} tables to answer (skipping any single table makes it unanswerable or semantically completely different).
        \item NOT mention the raw database column names explicitly (e.g., do not say "where artist\_id is 5"). Use natural language instead.
        \item Include specific proper nouns (names, places, IDs, events) from the data rows as anchors.
        \item NOT use generic phrasing like "connected to" or "in this table". Use precise verbs representing the table's schema role.
    \end{itemize}
    \item \{mid\_rule\}
    \item \textbf{Column Selection:} Target \{target\_c\} columns total across all \{n\_tables\} tables (Max \{max\_c\}).
\end{enumerate}

\vspace{0.8em}
\textbf{[EXPECTED JSON FORMAT]}
\begin{verbatim}
{
  "questions": [
    {
      "question": "<Natural language question reflecting the explicit JOIN relationships>",
      "thought": "{thought_tail}",
{col_lines}
    }
  ]
}
\end{verbatim}
\end{promptbox}
\caption{Prompt template used for explicit multi-hop query generation over structural FK/PK join chains.}
\end{figure*}
\begin{figure*}[t]
\small
\centering
\begin{promptbox}{Implicit 2-hop Join Prompt}
\label{fig:implicit_pair_prompt}
\ttfamily\setlength{\parindent}{0pt}

\textbf{[SYSTEM PROMPT]} \\
You are an expert in "Discriminative Multi-hop Retrieval". \\
Your goal is to generate ONE complex question that acts as a unique signature for the semantic connection between these TWO tables.

\vspace{0.5em}
\#\#\# CONTEXT: IMPLICIT (SEMANTIC) JOIN \\
These two tables are NOT connected by an explicit foreign key. Instead, they share overlapping real-world entities through semantically related columns. Think of it as: "Table 1 tells you one thing about entity X; Table 2 tells you another thing about entity X."

\vspace{0.5em}
\#\#\# STRATEGY: CROSS-PERSPECTIVE ENTITY ANCHORING
\begin{enumerate}[leftmargin=1.5em, noitemsep, topsep=2pt]
    \item \textbf{Find the Shared Entity:} Identify the entity appearing in BOTH tables under different column names.
    \item \textbf{Use Table 1 to Identify:} Pick a unique attribute from Table 1 to pinpoint an instance.
    \item \textbf{Retrieve from Table 2:} Ask for an attribute that only exists in Table 2.
    \item \textbf{Make It Discriminative:} Include enough specific proper nouns so only this pair can answer it.
\end{enumerate}

\vspace{0.5em}
\#\#\# RULES
\begin{enumerate}[leftmargin=1.5em, noitemsep, topsep=2pt]
    \item Output ONLY ONE valid JSON object:
\begin{verbatim}
{
  "questions": [
    {
      "question": "<Discriminative implicit multi-hop question>",
      "thought": "Step-by-step reasoning trace.",
      "relevant_columns_t1": ["col1"],
      "relevant_columns_t2": ["col2"]
    }
  ]
}
\end{verbatim}
    \item The question MUST:
    \begin{itemize}[leftmargin=1.em, noitemsep, topsep=1pt]
        \item Require BOTH tables to answer.
        \item NOT mention shared column names explicitly.
        \item Include specific proper nouns from data rows.
        \item NOT use pronouns or generic phrases ("in this table", "between the tables").
        \item Read like a simple factual lookup; the multi-table requirement must be inferrable only from data.
    \end{itemize}
    \item Output ONLY the JSON object. No markdown code fences. No extra text.
\end{enumerate}

\vspace{0.5em}
\#\#\# ABSOLUTE CONSTRAINT: EXACTLY 2 TABLES
\begin{itemize}[leftmargin=1.5em, noitemsep, topsep=2pt]
    \item Question MUST be answerable from EXACTLY these 2 tables.
    \item \textbf{Forbidden thought tokens:} "step 3", "table 3", "third table", "next look up in another".
    \item If a third table is required, output: \\
    \texttt{\{"questions":[\{"question":null,"thought":"requires more than 2 tables",...\}]\}}
\end{itemize}
\end{promptbox}
\caption{Prompt template for 2-hop implicit query generation over value-based semantic joins.}
\end{figure*}

\begin{figure*}[t]
\small
\centering
\begin{promptbox}{Implicit 3-hop Chain Prompt}
\label{fig:implicit_chain_prompt}
\ttfamily\setlength{\parindent}{0pt}

\textbf{[SYSTEM PROMPT]} \\
You are an expert in "Semantic Multi-hop Table Retrieval". \\
Your goal is to generate ONE question that acts as a unique signature for this 3-table semantic chain.

\vspace{0.5em}
\#\#\# CONTEXT: IMPLICIT SEMANTIC CHAIN (3 TABLES) \\
These 3 tables are connected by VALUE OVERLAP, NOT by FK/PK column-name patterns. \{bridge\_note\} \\
Table 2 is the bridge — it holds the entity that links Table 1 and Table 3 through shared values. Skipping Table 2 makes the question unanswerable or semantically different.

\vspace{0.5em}
\#\#\# STRATEGY: CHAINED ENTITY ANCHORING
\begin{enumerate}[leftmargin=1.5em, noitemsep, topsep=2pt]
    \item \textbf{Identify the bridge:} Table 2 connects Table 1 and Table 3 via shared values.
    \item \textbf{Anchor:} Use a specific proper noun from Table 1 (or Table 3) as the starting point.
    \item \textbf{Traverse:} Match through Table 2 first, then retrieve from the far end.
    \item \textbf{Discriminative:} Include specific proper nouns so only this 3-table chain can answer it.
\end{enumerate}

\vspace{0.5em}
\#\#\# RULES
\begin{enumerate}[leftmargin=1.5em, noitemsep, topsep=2pt]
    \item Output ONLY ONE valid JSON object:
\begin{verbatim}
{
  "questions": [
    {
      "question": "<Discriminative 3-table semantic chain question>",
      "thought": "1. Semantic Bridge: ... 2. Data Trace: ...
                  3. Indispensability: Table 2 ([name]) is
                  indispensable because [...]",
      "relevant_columns_t1": ["col1"],
      "relevant_columns_t2": ["col1"],
      "relevant_columns_t3": ["col1"]
    }
  ]
}
\end{verbatim}
    \item The question MUST:
    \begin{itemize}[leftmargin=1.em, noitemsep, topsep=1pt]
        \item Require ALL 3 tables to answer.
        \item NOT mention shared column names.
        \item Include specific proper nouns from data rows.
        \item NOT use pronouns.
        \item Read like a natural factual lookup; the 3-table requirement must not be hinted at.
    \end{itemize}
    \item Output ONLY the JSON object. No markdown code fences. No extra text.
\end{enumerate}
\end{promptbox}
\caption{Prompt template for 3-hop implicit query generation over semantic chain joins.}
\end{figure*}

\begin{figure*}[t]
\small
\centering
\begin{promptbox}{Single-hop Prompt}
\label{fig:single_prompt}
\ttfamily\setlength{\parindent}{0pt}

\textbf{[SYSTEM PROMPT]} \\
You are an expert in "Discriminative Table Retrieval".\\
Your goal is to generate 5 distinct questions that act as a \textbf{unique signature} for this table, ensuring it can be found among thousands of other similar tables.

\vspace{0.5em}
\#\#\# STRATEGY: DISCRIMINATIVE ANCHORING
\begin{enumerate}[leftmargin=1.5em, noitemsep, topsep=2pt]
    \item \textbf{Identify Primary Entities:} Find the main subjects (e.g., specific player names, specific city names, specific years, unique award titles).
    \item \textbf{Target Specific Rows:} Don't just summarize. Create questions that anchor to the most 'unique' or 'extreme' rows in the table.
    \item \textbf{Keyword Density:} Ensure every generated question contains the \textbf{exact proper nouns} (Names, Titles, Dates) found in the table.
\end{enumerate}

\vspace{0.5em}
\#\#\# RULES
\begin{enumerate}[leftmargin=1.5em, noitemsep, topsep=2pt]
    \item Output ONLY ONE valid JSON object following the exact schema provided in the format block below.
    \item Generate \textbf{exactly 5} questions with different levels of granularity:
    \begin{itemize}[leftmargin=1.0em, noitemsep]
        \item \textbf{Q1 (Entity-Focus):} Targets the main person/subject and their most notable record or identity.
        \item \textbf{Q2 (Temporal-Focus):} Links a specific year/season/date to a unique result in that period.
        \item \textbf{Q3 (Comparison/Superlative):} Focuses on "highest", "lowest", "first", or "total" values while mentioning the subject name.
        \item \textbf{Q4 (Multi-Column Integration):} Requires combining data from several different columns to form a complete answer.
        \item \textbf{Q5 (High-Complexity Reasoning):} A discriminative question that links multiple attributes to uniquely identify a specific data point.
    \end{itemize}
    \item \textbf{Question Constraints:} NEVER use generic phrases like "in this table" or "the provided data". ALWAYS include specific subject names. BE SPECIFIC. Ensure valid JSON format and escape backslashes properly (e.g., use \textbackslash\textbackslash\textbackslash\textbackslash\ inside strings).
    \item \textbf{Column Selection Guidelines:} Select the minimal set of relevant columns needed to answer the question to maintain optimal signal density, avoiding unnecessary information noise.
    \item Output ONLY the JSON object. No markdown code fences, no extra text.
\end{enumerate}

\vspace{0.8em}
\textbf{[EXPECTED OUTPUT FORMAT \& USER INPUT]}
\begin{verbatim}
{
  "feta_id": {table_id},
  "table_name": "{table_name}",
  "questions": [
    {
        "question": "<Discriminative, self-contained question containing specific proper nouns>",
        "thought": "<Step-by-step reasoning: Explain the logic of column composition>",
        "relevant_columns": ["col1", "col2"]
    }
  ]
}

### INPUT TABLE DATA
{table_str}

Generate the JSON object with 5 discriminative questions as specified.
\end{verbatim}
\end{promptbox}
\caption{Prompt template used for single-hop query generation.}
\end{figure*}


\textbf{Implicit Multi-Hop Query Prompt Template}
For implicit paths, we design separate prompts for 2-hop and 3-hop settings. The 2-hop prompt enforces a strict two-table setting with value-based entity alignment, while the 3-hop prompt treats the intermediate table as a value-based bridge connecting the end-point tables and adapts its description for semantic and mixed chains. Both prompts enforce multi-hop reasoning and require justification of each constituent table’s necessity.

\textbf{Explicit Multi-Hop Query Generation Prompt Template}
For explicit join paths, the prompt enforces alignment between semantic transitions and underlying Foreign-key relationships across 2-hop and 3-hop joins. Each generated query must require all tables in the chain to answer, and intermediate tables must be expressed as natural language semantic relations.

\textbf{Single-Hop Instructional Query Generation Prompt Template}
To extend \algname{} to single-hop retrieval, we design a prompt that generates diverse instructional queries over individual tables. As described in Section~\ref{app: quocca}, the queries cover a broad range of reasoning patterns, producing enriched table representations.

\end{document}